\documentclass[aps,prc,twocolumn,superscriptaddress,preprintnumbers,showpacs,floatfix,nofootinbib]{revtex4}

\usepackage{amsmath,graphicx}
\usepackage[colorlinks=true,linktocpage=true,linkcolor=blue,citecolor=blue]{hyperref}
\usepackage[usenames,dvipsnames]{color}
\usepackage{float}
\usepackage[normalem]{ulem}
\usepackage{comment}
\usepackage{textcomp}
\usepackage{bm}

\newcommand{\feq}{f_\mathrm{eq}}
\newcommand{\frs}{f_\mathrm{RS}}
\newcommand{\brs}{\beta_\mathrm{RS}}

\begin{document}

\title{On higher order and anisotropic hydrodynamics for Bjorken and Gubser flows}

\author{Chandrodoy Chattopadhyay}
\affiliation{Department of Nuclear and Atomic Physics, Tata Institute of Fundamental Research, Homi Bhabha Road, Mumbai 400005, India}
\author{Ulrich Heinz}
\affiliation{Department of Physics, The Ohio State University, Columbus, Ohio 43210, USA}
\affiliation{Theoretical Physics Department, CERN, CH-1211 Gen\`eve 23, Switzerland}
\author{Subrata Pal}
\affiliation{Department of Nuclear and Atomic Physics, Tata Institute of Fundamental Research, Homi Bhabha Road, Mumbai 400005, India}
\author{Gojko Vujanovic}
\affiliation{Department of Physics, The Ohio State University, Columbus, Ohio 43210, USA}
\preprint{CERN-TH-2018-005}
\date{\today}


\begin{abstract}
We study the evolution of hydrodynamic and non-hydrodynamic moments of the distribution function using anisotropic and third-order Chapman-Enskog hydrodynamics for systems undergoing Bjorken and Gubser flows. The hydrodynamic results are compared with the exact solution of the Boltzmann equation with a collision term in relaxation time approximation. While the evolution of the hydrodynamic moments of the distribution function (i.e. of the energy momentum tensor) can be described with high accuracy by both hydrodynamic approximation schemes,  their description of the evolution of the entropy of the system is much less precise. We attribute this to large contributions from non-hydrodynamic modes coupling into the entropy evolution which are not well captured by the hydrodynamic approximations. The differences between the exact solution and the hydrodynamic approximations are larger for the third-order Chapman-Enskog hydrodynamics than for anisotropic hydrodynamics, which effectively resums some of the dissipative effects from anisotropic expansion to all orders in the anisotropy, and are larger for Gubser flow than for Bjorken flow. Overall, anisotropic hydrodynamics provides the most precise macroscopic description for these highly anisotropically expanding systems. 
\end{abstract}

\maketitle

\section{Introduction}
\label{sec1}

A remarkable property of the hot and dense matter formed in ultra-relativistic heavy ion collisions at RHIC and LHC is a strong collective motion which has been successfully modeled using relativistic hydrodynamics (see \cite{Heinz:2013th} for a recent review). Dissipative hydrodynamics is formulated as an expansion in gradients of the fluid four-velocity, the simplest of them being the first-order Navier-Stokes theory due to Eckart \cite{Eckart:1940zz} and Landau and Lifshitz \cite{Landau}. Second-order dissipative theories developed later by Grad \cite{Grad}, M\"uller \cite{Muller:1967zza} and Israel and Stewart \cite{Israel:1979wp} cure an undesirable feature of relativistic Navier-Stokes theory, its acausality and instability \cite{Hiscock:1983zz,Hiscock:1985zz}. These theories, based on the principle of non-negative entropy production, are formulated by assuming an algebraic form for the entropy-four current in terms of dissipative quantities. Unfortunately, this method does not provide a unique set of higher-order viscous evolution equations. This has motivated a broad spectrum of attempts to derive dissipative relativistic hydrodynamics from a more fundamental framework.

Hydrodynamics may be regarded as a macroscopic effective theory of a many-body system in which the complex interactions occurring over short distance and time scales are averaged out, and the effective degrees of freedom are a small number of conserved charge currents coupled to dissipative fluxes. For sufficiently weak coupling among its microscopic constituents, such a system can be described statistically by a more involved kinetic theory, based on a single particle phase-space distribution function $f(x,p)$ whose evolution is typically governed by some generalized form of Boltzmann equation. The macroscopic conserved currents and dissipative fluxes can be formulated in terms of momentum moments of this distribution function for which equations of motion are then derived from the Boltzmann equation. Closing the set of moment equations requires approximations to truncate the resulting moment hierarchy. Different such approximation schemes result in different sets of hydrodynamic equations. The validity and accuracy of the applied approximations can be judged by comparing, for specific highly symmetric situations in which the underlying kinetic theory can be solved exactly, the solutions of the different hydrodynamic approximations to the corresponding momentum moments of the exact microscopic solution \cite{Florkowski:2013lya, Denicol:2014tha, Tinti:2015xra, Martinez:2017ibh}. This idea has generated increased interest for the search of new exact solutions of the relativistic Boltzmann equation \cite{Baym:1984np, Florkowski:2013lya, Florkowski:2014sfa, Denicol:2014xca, Florkowski:2014sda, Noronha:2015jia, Bazow:2015dha, Heinz:2015gka, Bazow:2016oky}.

The equilibrium distribution function in the local rest frame (LRF) of a fluid is, by definition, isotropic in the momentum space, irrespective of the macroscopic motion of the fluid. Using it as an approximation for the true LRF distribution function in rapidly expanding systems is justified only in the limit of vanishing mean free path, i.e. instantaneous local thermalization. For realistic systems with small but non-zero mean free paths this approximation fails to properly account for the competition between microscopic scattering processes driving the system towards local momentum isotropy (and eventually into local thermal equilibrium) and the macroscopic expansion rate which drives the local phase-space distribution away from local thermal equilibrium (and, in the case of anisotropic expansion, also away from local momentum isotropy). This leads to a deviation $\delta f(x,p)$ of the distribution function from its local equilibrium form, $f(x,p)=\feq(x,p)+\delta f(x,p)$, with the relative size of $\delta f$ increasing with the Knudsen number, i.e. with the product of the microscopic mean free time between collisions and the macroscopic expansion rate $\theta=\partial\cdot u$ where $u^\mu(x)$ is the fluid's flow four-velocity.

The first attempt to include such non-equilibrium $\delta f$ effects in the distribution function $f(x,p)$ was based on Grad's 14-moment approximation \cite{Grad, Israel:1979wp, Muronga:2003ta}. However, this moment expansion does not follow systematically from the underlying kinetic theory, such as the Boltzmann equation. A systematic approach of obtaining viscous hydrodynamics, to any given order in gradients of the macroscopic flow velocity, is based on a Chapman-Enskog-like iterative solution of Boltzmann equation \cite{Chapman, York:2008rr, Jaiswal:2012qm, Jaiswal:2013npa, Jaiswal:2013fc, Bhalerao:2013aha, Bhalerao:2013pza, Jaiswal:2013vta, Jaiswal:2014isa, Chattopadhyay:2014lya}. Recently, this method was employed to derive higher order dissipative hydrodynamic equations \cite{Jaiswal:2013vta}. Another novel way of formulating hydrodynamics from kinetic theory is based on an expansion controlled by the Knudsen number and the inverse Reynold's number \cite{Denicol:2012cn}. For a conformal system and using the relaxation time approximation (RTA) for the collision term of Boltzmann equation, this approach leads to identical viscous evolution equations as obtained in \cite{Jaiswal:2013vta}, up to second order in gradients.
 
All these formulations, however, assume that the local deviations of $f(x,p)$ from equilibrium are small, and that an expansion of $f(x,p)$ about its equilibrium value to a few low orders in derivatives should suffice. Anisotropic hydrodynamics \cite{Romatschke:2003ms, Martinez:2010sc, Ryblewski:2012rr, Bazow:2013ifa, Tinti:2013vba, Nopoush:2014qba, Tinti:2015xwa, Tinti:2015xra, Bazow:2016oky, Molnar:2016vvu} aims to extend the domain of applicability of traditional hydrodynamics, i.e., it attempts to better describe physical situations where the deviation of $f(x,p)$ from local momentum isotropy is non-perturbatively large. This is achieved by explicitly including in the leading-order LRF distribution function an anisotropy parameter $\xi$ describing the momentum-space deformation along the direction of largest anisotropy of the local expansion rate, and then expanding perturbatively the dynamical equations for the residual dissipative effects caused by the residual deviation $\delta\tilde f$, defined by writing $f(x,p)\equiv f_a(x,p;\xi)+\delta\tilde f(x,p)$. The non-trivial additional task in this approach is to determine the time evolution of the anisotropy parameter $\xi$ non-perturbatively such that the residual dissipative effects encoded in $\delta\tilde f$ are minimized and can again be described perturbatively. The recent works \cite{Tinti:2015xwa, Molnar:2016gwq, Martinez:2017ibh} have made significant progress in this direction.

It is necessary that the different macroscopic hydrodynamic formalisms described above are tested in scenarios where the microscopic dynamics can be solved exactly. We here study expanding systems with longitudinal boost-invariance and reflection symmetry, and either transverse homogeneity ((0+1)-dimensional Bjorken flow \cite{Bjorken:1982qr}) or azimuthally symmetric transverse density and flow gradients dictated by Gubser symmetry \cite{Gubser:2010ze} ((1+1)-dimensional Gubser flow \cite{Gubser:2010ze, Gubser:2010ui}). For these highly symmetric flow patterns in each case a convenient system of coordinates can be found in which the macroscopic hydrodynamic flow appears static and the microscopic relativistic Boltzmann equation, using the relaxation time approximation (RTA) for the collision term \cite{Anderson_Witting}, reduces to an ordinary differential equation in longitudinal proper time $\tau$ \cite{Bjorken:1982qr} or de Sitter time $\rho$ \cite{Gubser:2010ze}, respectively, and can be easily solved analytically \cite{Baym:1984np,Florkowski:2013lya,Denicol:2014tha}.

In this work, we compare with these exact solutions the evolution of various macroscopic variables obtained using hydrodynamic equations obtained from the (perturbative) third-order Chapman Enskog (CE) approach \cite{Jaiswal:2013vta,Chattopadhyay:2014lya} and the (non-perturbative) anisotropic hydrodynamic approach in the $P_L$ matching scheme \cite{Molnar:2016gwq, Martinez:2017ibh}. The present work goes beyond similar earlier comparisons \cite{Denicol:2014tha, Chattopadhyay:2014lya, Molnar:2016gwq, Martinez:2017ibh} by presenting for the first time the solution of third-order CE evolution equations for Gubser flow and a detailed analysis of the evolution of the systems'\ entropy in the various approximations (see \cite{Bazow:2016oky} for an earlier study of entropy production in the isotropic FLRW universe). We find that entropy production is a sensitive discriminator between different hydrodynamic approximations and exhibits generically much larger deviations from the exact solution of the Boltzmann equation than all of the hydrodynamic observables. This reflects a significant contribution to entropy production by non-hydrodynamic modes whose dynamics is not constrained by macroscopic conservation laws.

The paper is organized as follows. In Section~\ref{sec2} we briefly describe the Bjorken and Gubser flow profiles and the coordinates we use to describe them. Section~\ref{sec3} reviews the exact solution of the Boltzmann equation in relaxation time approximation for the two flow profiles. In Sec.~\ref{sec4} we elaborate on the Chapman-Enskog formalism and derive third-order dissipative hydrodynamics for Gubser flow. This is followed in Sec.~\ref{sec5} by a brief review of anisotropic hydrodynamics in the $P_L$ matching scheme for the Bjorken and Gubser flows. Numerical results for the comparison of the different approaches are presented and discussed in Sec.~\ref{sec6}. We close with conclusions and an outlook in Sec~\ref{sec7}.

\section{Bjorken and Gubser flows}
\label{sec2}

Bjorken flow \cite{Bjorken:1982qr} is most naturally expressed in Milne coordinates $(\tau,r,\phi,\eta)$,
\begin{align}
\nonumber
 \tau &= \sqrt{t^2 - z^2}, \quad \quad \eta = \tanh^{-1}\left( \frac{z}{t} \right), 
 \\\label{eq1}
 r &= \sqrt{x^2 +y^2}, \quad \quad \phi = \tan^{-1}\left( \frac{y}{x} \right),
\end{align}
with the metric $g_{\mu\nu} = \mathrm{diag}(-1,1,r^2,\tau^2)$ (in ``mostly plus'' convention) and line element
\begin{align}
\label{eq2}
 ds^2 = -d\tau^2 + dr^2 + r^2 d\phi^2 + \tau^2 d\eta^2.
\end{align}
Equation (\ref{eq2}) is manifestly invariant under the Bjorken symmetry, namely boost-invariance  ($SO(1,1)$) along the beam direction $\eta$, rotational and translational invariance in the transverse $(x,y)$ plane ($ISO(2)$), and reflection ($Z_2$) symmetry under $\eta\to -\eta$. The only flow consistent with the combined $ISO(2)\otimes SO(1,1) \otimes Z_2$ symmetry group is $u^{\mu} \equiv (u^\tau,u^x,u^y,u^\eta) = (1,0,0,0)$, in association with 
$(r,\phi,\eta)$ independence of all macroscopic physical quantities.

Gubser \cite{Gubser:2010ze} relaxed the $ISO(2)$ symmetry of Bjorken flow, replacing it with symmetry under the $SO(3)_q$ (conformal) group of transformations, while maintaining the invariance under boosts and reflections. Gubser flow appears static in de Sitter coordinates on a curved spacetime formed by the direct product of a three-dimensional de Sitter space ($dS_3$) with a line, $dS_3 \otimes R$, defined by a Weyl rescaling of the metric in Milne coordinates,
\begin{align}
\label{eq3}
 d\hat{s}^2 = \frac{ds^2}{\tau^2}= \frac{-d\tau^2 +dr^2 +r^2d\phi^2}{\tau^2} + d\eta^2,
\end{align}
followed by a coordinate transformation to $\hat{x}^{\mu} = (\rho,\theta,\phi,\eta)$ where \cite{Gubser:2010ze}
\begin{align} 
\label{rhotheta}
 \rho &= - \sinh^{-1} \left( \frac{1-q^2\tau^2 + q^2 r^2}{2q\tau} \right), 
 \nonumber\\
 \theta &= \tan^{-1} \left( \frac{2qr}{1+q^2\tau^2-q^2 r^2}   \right).
\end{align}
Here $q$ is an arbitrary energy scale which sets the transverse size of the system. In these coordinates, the Weyl rescaled line element 
\begin{align} 
\label{linedS}
  d\hat{s}^2 = -d\rho^2 + \cosh^2 \rho (d\theta^2 + \sin^2\theta d\phi^2)  + d\eta^2,
\end{align}
with metric $\hat{g}_{\mu\nu} = \mathrm{diag} (-1,\cosh^2\rho,\cosh^2\rho\sin^2\theta,1)$, is manifestly symmetric under the Gubser group $SO(3)_q \otimes SO(1,1) \otimes Z_2$ since the $SO(3)_q$ conformal symmetry corresponds to standard ``rotations'' of the sphere parametrized by $(\theta,\phi)$. The Gubser flow becomes static in de Sitter coordinates, $\hat{u}^{\mu} \equiv (\hat{u}^\rho,\hat{u}^\theta,\hat{u}^\phi,\hat{u}^\eta)= (1,0,0,0)$, and all macroscopic variables depend only on the ``de Sitter time'' $\rho$.

In this paper, any quantity expressed in Gubser coordinates $\hat{x}^{\mu}$ is denoted by a hat and made unitless by scaling with appropriate powers of the Weyl rescaling parameter (longitudinal proper time in Milne coordinates) $\tau$ \cite{Gubser:2010ze,Gubser:2010ui}. For example,
\begin{align}
\label{eq6}
  \epsilon(\tau,r) = \frac{\hat{\epsilon}(\rho)}{\tau^4}, \qquad
  \pi_{\mu\nu}(\tau,r) =  \frac{1}{\tau^2} \frac{\partial \hat{x}^\alpha}{\partial x^{\mu}}
  \frac{ \partial \hat{x}^\beta}{\partial x ^{\nu}} \hat{\pi}_{\alpha \beta}(\rho).
\end{align}
%

\section{Exact solution of the Boltzmann equation for Bjorken and Gubser flows}
\label{sec3}

In this section, we review the central idea common to deriving from microscopic dynamics the dissipative hydrodynamic equations considered in this article. We consider a conformally symmetric system of weakly interacting massless Boltzmann particles without conserved charges whose phase-space distribution function $f(x,p)$ evolves according to the Boltzmann equation. In the absence of external forces, and with a relaxation-time approximation for the collisional kernel, the Boltzmann equation has the form \cite{Anderson_Witting}
\begin{equation}
\label{RBERTA}
   p^\mu\partial_\mu f  =   (u\cdot p)\frac{\delta f}{\tau_r},
\end{equation}
where $\tau_r(x)$ is the momentum-independent relaxation time and $\delta f\equiv f-\feq$ is the deviation of the distribution function from its local equilibrium form $\feq(x,p) \equiv \exp[-\beta(x)\, (p \cdot u(x))]$. Here $\beta(x) \equiv 1/T(x)$ is the inverse local temperature and $u^{\mu}(x)$ is the velocity of the local rest frame, defined as the velocity associated with the local energy flow (LRF = Landau frame). Conformal symmetry requires $\tau_r = 5 \bar{\eta}/T \equiv c/T$, where specific shear viscosity $\bar{\eta}\equiv \eta/s$ is defined the ratio of shear viscosity $\eta$ to entropy density $s$.

On obtaining a solution of Eq.~(\ref{RBERTA}), either exact or in some approximation, the macroscopic hydrodynamic variables are constructed from the momentum moments of $f(x,p)$. Specifically, the conserved energy momentum tensor $T^{\mu\nu}$ is the second moment of 
$f(x,p)$ \cite{deGroot}:
\begin{align} 
\label{EMRTA}
 T^{\mu\nu} \equiv \langle p^{\mu} p^{\nu} \rangle,
\end{align}
where we use the shorthand notation $\langle {\cal O}(x)  \rangle \equiv \int dp \, {\cal O}(x,p) f(x,p) $ where
$dp \equiv d^3p/[(2 \pi)^3|\bm{p}|\sqrt{-g}]$ is the invariant momentum-space integration measure,
with $g$ being the determinant of the metric tensor.

In the remainder of this section we discuss the exact solutions of Eq.~(\ref{RBERTA}) for Bjorken and Gubser flows. Two other, more general methods of obtaining from Eq.~(\ref{RBERTA}) approximate solutions for $f(x,p)$, namely the Chapman-Enskog iterative scheme and anisotropic hydrodynamics with $P_L$ matching, are presented in the next two sections, including again their specific forms for Bjorken and Gubser flows. 

For massless systems with Bjorken symmetry the single particle phase-space distribution $f(x,p) = f(\tau;p_T,w)$ can only depend on the longitudinal proper time $\tau$, the magnitude of the transverse momentum $p_T$, and the longitudinally boost-invariant variable $w = t p^z - z p^0 = p_T\tau\sinh(y{-}\eta)$ (where $y$ is the kinematic rapidity of a particle) \cite{Baym:1984np,Florkowski:2013lya}. With this simplification, Eq.~(\ref{RBERTA}) reduces, at every point $(p_T,w)$ in momentum space, to an ordinary differential equation in $\tau$,
\begin{align}
 \frac{\partial f}{\partial \tau} = - \frac{f{-}\feq}{\tau_r},
\end{align}
with the integral solution \cite{Florkowski:2013lya}
\begin{align}
\label{Bj_soln}
  f(\tau;p_T,w) =& D(\tau,\tau_0)f(\tau_0;p_T,w) 
\nonumber \\
  +& \int_{\tau_0}^{\tau} \frac{d\tau'}{\tau_r(\tau')} D(\tau,\tau') \feq\bigl(p^\tau(\tau')/T(\tau')\bigr).
\end{align}
Here $D(\tau_2,\tau_1) = \exp\bigl(-\int_{\tau_1}^{\tau_2} d\tau ' /\tau_r(\tau')\bigr)$ is the so-called damping function, and the energy $p^\tau$ is obtained from $p_T$ and $w$ through the mass-shell constraint. The temperature defining the local equilibrium distribution under the integral in the last term is obtained from the Landau matching condition $\epsilon =  \langle (u\cdot p)^2\rangle = \epsilon_\mathrm{eq} = \langle (u\cdot p)^2\rangle_\mathrm{eq} = (3/\pi^2) T^4$. This condition involves an integral over all momenta $(p_T,w)$ and renders the solution (\ref{Bj_soln}) highly nonlinear, in spite of its apparent simplicity.

For Gubser flow the symmetries constrain the dependence of the phase-space distribution $f$ as follows: $f(\hat{x},\hat{p}) = f(\rho;\hat{p}^2_\Omega, \hat{p}_\eta)$. Here $\hat{p}^2_\Omega = \hat{p}^2_\theta + \hat{p}^2_\phi/\sin^2\theta$ plays the role of transverse momentum \cite{Denicol:2014xca,Denicol:2014tha}, and $\hat{p}_\eta=w$ is the same boost-invariant longitudinal momentum variable as in the Bjorken case. Again, the symmetry constraints reduce the RTA Boltzmann equation (\ref{RBERTA}) to an ordinary differential equation at each point $(\hat{p}^2_\Omega,\hat{p}_\eta)$ in momentum space,
\begin{align}
   \frac{\partial f(\rho;\hat{p}^2_\Omega,\hat{p}_\eta)}{\partial \rho} = 
          -\frac{\hat{T}(\rho)}{c} \left( f(\rho;\hat{p}^2_\Omega,\hat{p}_\eta) - \feq(\hat{p}^\rho/\hat{T}(\rho)) \right),
\end{align}
with the solution \cite{Denicol:2014tha,Denicol:2014xca}
\begin{align}
\label{G_soln}
 f(\rho;\hat{p}_{\Omega}^2,\hat{p}_{\eta}) =& D(\rho,\rho_0) f(\rho_0;\hat{p}_{\Omega}^2,\hat{p}_{\eta})
 \\\nonumber
 +&\frac{1}{c}\int_{\rho_0}^{\rho} d\, \rho' D(\rho,\rho') \hat{T}(\rho')
      \feq\bigl(\hat{p}^\rho(\rho')/\hat{T}(\rho')\bigr)
\end{align}
where $D(\rho_2,\rho_1) = \exp\bigl( - \int_{\rho_1}^{\rho_2} d\rho' \, \hat{T}(\rho')/c \bigr)$. Again, the temperature in the equilibrium distribution on the right hand side is obtained by Landau matching, and $\hat{p}^\rho$ is obtained from $(\hat{p}^2_\Omega,\hat{p}_\eta)$ through the mass-shell constraint.

The exact solutions (\ref{Bj_soln},\ref{G_soln}) can be evaluated numerically \cite{Florkowski:2013lya,Denicol:2014tha}, and the exact evolution of any macroscopic quantity (in particular of all the components of the energy momentum tensor (\ref{EMRTA})) can then be obtained by taking appropriate momentum moments of the exact $f(x,p)$.

\section{Dissipative hydrodynamics from the Chapman-Enskog method}
\label{sec4}

This method is based on the assumption that the deviation of $f(x,p)$ from its local equilibrium value is small, such that the RTA Boltzmann equation, Eq. (\ref{RBERTA}), can be solved iteratively to obtain a Chapman-Enskog-like expansion for the non-equilibrium part of the distribution function in powers of space-time gradients \cite{Chapman,Romatschke:2011qp}:
\begin{equation}
\label{CEE} 
   \delta f = \delta f^{(1)} + \delta f^{(2)} + \delta f^{(3)} + \cdots, 
\end{equation}
where $\delta f^{(1)}$ is first-order in derivatives, $\delta f^{(2)}$ is second-order, and so on. To first and second order in derivatives one obtains
\begin{align}
\label{FOC}
   \delta f^{(1)} &= \frac{\tau_r}{u\!\cdot\! p} \, p^\mu \partial_\mu \feq,   
\\\label{SOC}
   \delta f^{(2)} &= \frac{\tau_r}{u\!\cdot\! p}p^\mu p^\nu\partial_\mu\Big(\frac{\tau_r}{u\!\cdot\! p} \partial_\nu \feq\Big). 
\end{align}
The above expansion may also be seen as a perturbation series in powers of the expansion parameter $\tau_r$. 

The energy-momentum tensor has the general form
\begin{align}
\label{NTD}
  T^{\mu\nu} &=  \langle p^\mu p^\nu\rangle =\epsilon u^\mu u^\nu + P\Delta ^{\mu \nu} +  \pi^{\mu\nu},  
\end{align}
where $\epsilon$ and $P$ are local energy density and pressure, respectively, related to the temperature by $\epsilon = 3P = 3/(\pi^2\beta^4)$ through the Landau matching condition. $\Delta^{\mu\nu} \equiv g^{\mu\nu} + u^{\mu}u^{\nu} $ projects a tensor to the space orthogonal to $u^{\mu}$, and the shear stress tensor $\pi^{\mu\nu}$ is traceless and orthogonal to $u^{\mu}$. 

The evolution equations for $\epsilon$ and $u^\mu$ are obtained from energy-momentum conservation, $\partial_\mu 
T^{\mu\nu} =0$:
\begin{align}
\label{evol01}
  \dot\epsilon + (\epsilon+P)\theta + \pi^{\mu\nu}\sigma_{\mu\nu} &= 0,  
\\
\label{evol02}
  (\epsilon+P)\dot u^\alpha + \nabla^\alpha P + \Delta^\alpha_\nu \partial_\mu \pi^{\mu\nu}  &= 0. 
\end{align}
We use the standard notation $\dot A\equiv u^\mu\partial_\mu A$ for the co-moving time derivative, $\theta\equiv \partial_\mu u^\mu$ for the expansion scalar, $\sigma_{\mu\nu}\equiv(\nabla_{\mu}u_{\nu} + \nabla_{\nu}u_{\mu})/2-(\theta/3)\Delta_{\mu\nu}$ for the velocity shear tensor, and $\nabla^\alpha\equiv\Delta^{\mu\alpha} 
\partial_\mu$ for space-like derivatives in the LRF.

To close the equations (\ref{evol01},\ref{evol02}) we need additional equations for the shear stress $\pi^{\mu\nu}$. To obtain them we express $\pi^{\mu\nu}$ in terms of $\delta f$,
\begin{align}
\label{FSE}
   \pi^{\mu\nu} &= \Delta^{\mu\nu}_{\alpha\beta} \int dp \, p^\alpha p^\beta\, \delta f,
\end{align}
where $\Delta^{\mu\nu}_{\alpha\beta} \equiv \Delta^{\mu}_{(\alpha}\Delta^{\nu}_{\beta)} - (1/3)\Delta^{\mu\nu}\Delta_{\alpha\beta}$ is a traceless symmetric projection operator orthogonal to $u^\mu$, with $ \Delta^{\mu}_{(\alpha}\Delta^{\nu}_{\beta)} \equiv \frac{1}{2}\left(\Delta^\mu_\alpha \Delta^\mu_\beta + \Delta^\mu_\beta \Delta^\mu_\alpha \right)$. If one substitutes on the r.h.s. for $\delta f$ the first-order term (\ref{FOC}) of the expansion (\ref{CEE}) and uses the energy-momentum conservation laws (\ref{evol01}-\ref{evol02}) together with $\epsilon \propto \beta^{-4}$ to eliminate all temperature derivatives on the r.h.s. of Eq.~(\ref{FOC}) in terms of velocity gradients, one obtains the well-know Navier-Stokes result $\pi^{\mu\nu} = -2\tau_r \beta_\pi \sigma ^{\mu\nu} $. Here $\beta_\pi$ is a thermodynamic integral over the local equilibrium distribution, related to the relaxation time $\tau_r$ and shear viscosity $\eta$ by $\tau_r = \eta/\beta_\pi$.

To obtain higher order approximations for $\pi^{\mu\nu}$ we take the co-moving time derivative of Eq.~(\ref{FSE}),
\begin{equation}
\label{SSE}
  \dot\pi^{\langle\mu\nu\rangle} = \Delta^{\mu\nu}_{\alpha\beta} \int dp\, p^\alpha p^\beta\, \delta\dot f, 
\end{equation}
where $A^{\langle\mu\nu\rangle}\equiv \Delta^{\mu\nu}_{\alpha\beta}A^{\alpha\beta}$ denotes the traceless symmetric projection orthogonal to $u^{\mu}$ of the tensor $A^{\mu\nu}$, and express $\delta \dot{f}$ through the Boltzmann equation (\ref{RBERTA}), by rewriting it as
\begin{equation}
\label{DFD}
   \delta\dot f = -\dot{f}_\mathrm{eq} + \frac{1}{u\!\cdot\! p}p^\gamma\nabla_\gamma f - \frac{\delta f}{\tau_r}. 
\end{equation}
Inserting this back into Eq.~(\ref{SSE}) one obtains
\begin{equation}
\label{SOSE}
 \dot\pi^{\langle\mu\nu\rangle} + \frac{\pi^{\mu\nu}}{\tau_r} = 
 \Delta^{\mu\nu}_{\alpha\beta} \!\int\! \frac{dp}{u\!\cdot\! p} \, p^\alpha p^\beta p^\gamma\nabla_\gamma f. 
\end{equation}
From this equation it is clear that the shear relaxation time $\tau_\pi$ is equal to the Boltzmann relaxation time $\tau_r$. Now we can substitute $f=\feq+\delta f^{(1)}$, with $\delta f^{(1)}$ from Eq.~(\ref{FOC}), on the r.h.s. of Eq.~(\ref{SOSE}) to obtain the second-order evolution equation \cite{Jaiswal:2013npa} (see also \cite{Denicol:2012cn})
\begin{equation}\label{SOSHEAR}
  \dot{\pi}^{\langle\mu\nu\rangle} \!+ \frac{\pi^{\mu\nu}}{\tau_\pi}\!= 
  -2\beta_{\pi}\sigma^{\mu\nu} 
  \!+2\pi_\gamma^{\langle\mu}\omega^{\nu\rangle\gamma}
  \!-\frac{10}{7}\pi_\gamma^{\langle\mu}\sigma^{\nu\rangle\gamma} 
  \!-\frac{4}{3}\pi^{\mu\nu}\theta,
\end{equation}
where $\omega^{\mu\nu}\equiv \nabla^{[\mu}u^{\nu]} \equiv \frac{1}{2}(\nabla^\mu u^\nu{-}\nabla^\nu u^\mu)$ is the vorticity tensor. 

To go to third-order, $\delta f$ is required up to second-order in velocity gradients,
\begin{equation}\label{SOVC}
  \delta f = f_0\phi = \feq\left(\phi_1 + \phi_2\right) + {\cal O}(\delta^3),
\end{equation}
where $\phi_1$ and $\phi_2$ are first- and second-order corrections, respectively. They are found to be \cite{Jaiswal:2013vta}
\begin{align}
\label{phi1}
  \phi_1 =\, & -\frac{\beta}{2\beta_\pi(u\!\cdot\!p)}\, p^\alpha p^\beta \pi_{\alpha\beta},  
\\ 
\label{phi2} 
  \phi_2 =\, & \frac{\beta}{\beta_\pi} \bigg[\frac{5}{14\beta_\pi (u\!\cdot\!p)}\, p^\alpha p^\beta \pi^\gamma_\alpha\,
                     \pi_{\beta\gamma} + \frac{\tau_\pi}{u\!\cdot\!p}\, p^\alpha p^\beta \pi^\gamma_\alpha\,
                     \omega_{\beta\gamma} 
\\ 
   &+\frac{(u\!\cdot\!p)}{70\beta_\pi}\, \pi^{\alpha\beta}\pi_{\alpha\beta} 
   +\frac{6\tau_\pi}{5}\, p^\alpha\dot u^\beta\pi_{\alpha\beta}
   +\frac{\tau_\pi}{5}\, p^\alpha\!\left(\nabla^\beta\pi_{\alpha\beta}\!\right)  
\nonumber\\ 
   &-\frac{\tau_\pi}{2(u\!\cdot\!p)^2}\, p^\alpha p^\beta p^\gamma\!\left(\nabla_\gamma\pi_{\alpha\beta}\!\right) 
   -\frac{3\tau_\pi}{(u\!\cdot\!p)^2}\, p^\alpha p^\beta p^\gamma \pi_{\alpha\beta}\dot u_\gamma 
\nonumber\\ 
\nonumber
   &+\frac{\tau_\pi}{3(u\!\cdot\!p)}\, p^\alpha p^\beta \pi_{\alpha\beta}\theta 
   +\frac{\beta-(u\!\cdot\!p)^{-1}}{4(u\!\cdot\!p)^2\beta_\pi}\left(p^\alpha p^\beta
      \pi_{\alpha\beta}\right)^2\bigg]. 
\end{align}
Please note the change of sign of several terms in the above equation compared to \cite{Jaiswal:2013vta} where a ``mostly minus'' signature for the metric was used. We note that $\phi_1$ and $\phi_2$ in Eqs.~(\ref{phi1},\ref{phi2}) satisfy the Landau matching conditions $u_\nu T^{\mu \nu} = \epsilon u^\mu$ and $\epsilon = \epsilon_\mathrm{eq}$ \cite{Bhalerao:2013pza}. 

Substituting $f{\,=\,}\feq(1+\phi_1+\phi_2)$ into Eq.~(\ref{SOSE}), some algebra yields the third-order evolution equation \cite{Jaiswal:2013vta}
\begin{align}
\label{TOSHEAR}
  \dot{\pi}^{\langle\mu\nu\rangle} 
 =& -\frac{\pi^{\mu\nu}}{\tau_\pi} - 2\beta_\pi\sigma^{\mu\nu} 
  +2\pi_{\gamma}^{\langle\mu}\omega^{\nu\rangle\gamma}
  -\frac{10}{7}\pi_\gamma^{\langle\mu}\sigma^{\nu\rangle\gamma}  
\nonumber\\
   &-\frac{4}{3}\pi^{\mu\nu}\theta
   -\frac{25}{7\beta_\pi}\pi^{\rho\langle\mu}\omega^{\nu\rangle\gamma}\pi_{\rho\gamma} 
   +\frac{1}{3\beta_\pi}\pi_\gamma^{\langle\mu}\pi^{\nu\rangle\gamma} \theta 
\nonumber\\ 
   &+\frac{38}{245\beta_\pi}\pi^{\mu\nu}\pi^{\rho\gamma}\sigma_{\rho\gamma} 
   +\frac{22}{49\beta_\pi}\pi^{\rho\langle\mu}\pi^{\nu\rangle\gamma}\sigma_{\rho\gamma} 
\nonumber\\ 
   &-\frac{24}{35}\nabla^{\langle\mu}\left(\pi^{\nu\rangle\gamma}\dot u_\gamma\tau_\pi\right)
   -\frac{4}{35}\nabla^{\langle\mu}\left(\tau_\pi\nabla_\gamma\pi^{\nu\rangle\gamma}\right) 
\nonumber\\ 
   &+\frac{2}{7}\nabla_{\gamma}\left(\tau_\pi\nabla^{\langle\mu}\pi^{\nu\rangle\gamma}\right) 
   +\frac{12}{7}\nabla_{\gamma}\left(\tau_\pi\dot u^{\langle\mu}\pi^{\nu\rangle\gamma}\right) 
\nonumber\\
   &+\frac{1}{7}\nabla_{\gamma}\left(\tau_\pi\nabla^{\gamma}\pi^{\langle\mu\nu\rangle}\right) 
   +\frac{6}{7}\nabla_{\gamma}\left(\tau_\pi\dot u^{\gamma}\pi^{\langle\mu\nu\rangle}\right) 
\nonumber\\
   &-\frac{2}{7}\tau_\pi\omega^{\rho\langle\mu}\omega^{\nu\rangle\gamma}\pi_{\rho\gamma}
   -\frac{2}{7}\tau_\pi\pi^{\rho\langle\mu}\omega^{\nu\rangle\gamma}\omega_{\rho\gamma} 
\nonumber\\
   &-\frac{10}{63}\tau_\pi\pi^{\mu\nu}\theta^2 + \frac{26}{21}\tau_\pi 
       \pi_\gamma^{\langle\mu} \omega^{\nu\rangle\gamma} \theta.
\end{align}
The right-hand side of this equation contains three second-order and fourteen third-order terms. 

An expression for the entropy four-current is derived using the kinetic theory definition for particles with Boltzmann statistics \cite{deGroot}
\begin{equation}
\label{EFC}
    S^\mu = -\int dp ~p^\mu f \bigl(\ln f{-}1\bigr).
\end{equation}
Assuming small deviations from local thermodynamic equilibrium, 
$f=\feq(1+\phi)$, where $\phi\ll 1$, we obtain an expression for the 
non-equilibrium entropy four-current up to third-order in $\phi$ as
\begin{equation}
\label{TEFC}
   S^\mu = s_\mathrm{eq} u^\mu - \int dp ~p^\mu \feq \left(\frac{\phi^2}{2}-\frac{\phi^3}{6}\right),
\end{equation}
where $s_\mathrm{eq}=\beta(\epsilon+P)$ is the equilibrium definition of the entropy density. For $\phi=\phi_1+\phi_2$ we have
\begin{equation}
\label{TOEFC}
   S^\mu = s_\mathrm{eq} u^\mu - \int dp ~p^\mu \feq \left(\frac{\phi_1^2}{2}+\phi_1\phi_2-\frac{\phi_1^3}{6}\right),
\end{equation}
where we ignore terms higher than third-order in the derivative expansion. Substituting $\phi_1$ and $\phi_2$ from Eqs.~(\ref{phi1}) and (\ref{phi2}) and performing the integrations, we obtain \cite{Chattopadhyay:2014lya}
\begin{align}
\label{TOEFCF}
   S^\mu =&~ s_\mathrm{eq} u^\mu - \frac{\beta}{4\beta_\pi}\pi^{\alpha\beta}\pi_{\alpha\beta}u^\mu
   +\frac{5\beta}{42\beta_\pi^2}\pi_{\alpha\gamma}\pi^\gamma_\beta\pi^{\alpha\beta}u^\mu 
\nonumber\\ 
  &~+\!\frac{\beta\tau_\pi}{7\beta_\pi}\bigg[\frac{18}{5}\dot u^\rho \pi_{\rho\gamma}\pi^{\mu\gamma} 
   \!-\frac{2}{5}\pi^{\mu\gamma}\nabla^\rho\pi_{\rho\gamma} 
   \!+\frac{1}{2}\pi^{\alpha\beta}\nabla^\mu\pi_{\alpha\beta} 
\nonumber\\ 
   &~+3\dot u^\mu\pi_{\alpha\beta}\pi^{\alpha\beta}
   +\pi^{\alpha\gamma}\Delta^{\mu\rho}\nabla_\alpha\pi_{\rho\gamma}\bigg], 
\end{align}
recalling that $\beta_\pi=4P/5$. The LRF entropy density, $s\equiv - u_\mu S^\mu$, 
is given by
\begin{align}
\label{TOEDCE} 
   s =&~ s_\mathrm{eq} - \frac{\beta}{4\beta_\pi}\pi^{\alpha\beta}\pi_{\alpha\beta}
   +\frac{5\beta}{42\beta_\pi^2}\pi_{\alpha\gamma}\pi^\gamma_\beta\pi^{\alpha\beta}, 
\end{align}
whereas the entropy flux in the LRF, $S^{\langle\mu\rangle}\equiv\Delta^\mu_\nu S^\nu$, reduces to
\begin{align}
\label{TOEFLCE}
   S^{\langle\mu\rangle} =&~ \frac{\beta\tau_\pi}{7\beta_\pi}\bigg[\frac{18}{5}\dot u^\rho 
                                              \pi_{\rho\gamma}\pi^{\mu\gamma} 
   \!-\frac{2}{5}\pi^{\mu\gamma}\nabla^\rho\pi_{\rho\gamma} 
   \!+\frac{1}{2}\pi^{\alpha\beta}\nabla^\mu\pi_{\alpha\beta} 
\nonumber\\ 
   &\qquad\  +3\dot u^\mu\pi_{\alpha\beta}\pi^{\alpha\beta}
   + \pi^{\alpha\gamma}\Delta^{\mu\rho}\nabla_\alpha\pi_{\rho\gamma}\bigg]. 
\end{align}
We observe that, beginning at third order in the derivative expansion, the Chapman-Enskog method leads to a non-vanishing entropy flux in the LRF.

\subsection{Evolution equations in Bjorken flow}
\label{sec4a}

In this and the following subsection we simplify the hydrodynamic evolution equations (\ref{evol01},\ref{evol02}) and the evolution equation for the shear stress (\ref{TOSHEAR}) for Bjorken- and Gubser-symmetric systems, respectively. For Bjorken flow \cite{Bjorken:1982qr} we can follow \cite{Jaiswal:2013vta}. We observe that Bjorken symmetry implies $\omega^{\mu\nu}{\,=\,}\dot u^\mu{\,=\,}\nabla^\mu\tau_\pi{\,=\,}0$, $\theta{\,=\,}1/\tau$, $\sigma^{\eta\eta}{\,=\,}2/(3\tau^3)$, and that only the $\eta\eta$ component of Eq.~(\ref{TOSHEAR}) survives, which we write in terms of $\pi{\,\equiv\,}{-}\tau^2\pi^{\eta\eta}$. With these simplifications Eqs.~(\ref{evol01},\ref{evol02},\ref{TOSHEAR}) become
\begin{align}
\label{BED} 
  \frac{d\epsilon}{d\tau} &= -\frac{1}{\tau}\left(\frac{4}{3}\epsilon -\pi\right), 
\\
\label{Bshear}
   \frac{d\pi}{d\tau} &= - \frac{\pi}{\tau_\pi} 
   + \frac{1}{\tau}\left(\frac{4}{3}\beta_\pi - \lambda\pi - \chi\frac{\pi^2}{\beta_\pi}\right). 
\end{align}
In the last equation the terms proportional to $\lambda$ and $\chi$ are the only surviving second- and third-order terms, respectively. In order to rewrite some of the third-order contributions in the form $\pi^2/(\beta_\pi\tau)$, the first-order (Navier-Stokes) expression for the shear pressure, $\pi=(4/3)\beta_\pi\tau_\pi/\tau$, has 
been used. The transport coefficients in Eq.~(\ref{Bshear}) are simply
\begin{equation}\label{BTC}
  \beta_\pi = \frac{4P}{5}, \quad \lambda = \frac{38}{21}, \quad \chi = \frac{72}{245}.
\end{equation}
For Bjorken flow the entropy flux in the LRF vanishes, $S^{\langle \mu \rangle} = 0$, and the LRF entropy density can be written as
\begin{align}
\label{entrBJ}
  s(\tau) = s_\mathrm{eq} -\frac{3\beta}{8 \beta_\pi} \pi^2 - \frac{15\beta}{168\beta_\pi^2} \pi^3.
\end{align}

\subsection{Evolution equations in Gubser flow}
\label{sec4b}

For systems with Gubser symmetry $\hat{\pi}^{\mu\nu}$ is diagonal in de Sitter coordinates, with $\hat{\pi}^{\rho\rho} = 0$, and the shear stress tensor has only one independent component which we take as $\hat\pi^{\eta\eta}$: $ \hat{\pi}^{\theta}_{\theta} =  \hat{\pi}^{\phi}_{\phi} = - (\hat{\pi}^{\eta}_{\eta})/2 \equiv -\hat{\pi}/2$. Similar to Bjorken flow, the vorticity is zero, $\omega^{\mu\nu}=0$, and since the flow is static in de Sitter coordinates, the acceleration $\dot{u^{\mu}}$ vanishes. Furthermore, $\hat{\tau}_\pi\sim \hat\beta=1/\hat{T}$ depends only on the de Sitter time $\rho$, so $\hat{\nabla}^{\mu} \tau_{\pi} = 0$. With these simplifications the non-trivial terms in the $\eta\eta$ component of the shear stress evolution equation are
\begin{align}
\label{identity}
   \dot{\hat{\pi}}^{\langle\eta\eta\rangle}  &= \frac{d\hat{\pi}}{d\rho},  &
   \hat{\pi}^{\langle\eta}_{\gamma}\hat{\sigma}^{\eta\rangle\gamma} &= -\frac{\hat{\theta}}{6} \hat{\pi}, 
\nonumber\\
   \hat{\pi}^{\langle\eta}_{\gamma}\hat{\pi}^{\eta\rangle\gamma} &= \frac{\hat{\pi}^2}{2},  &
   \hat{\pi}^{\rho\gamma}\hat{\sigma}_{\rho\gamma} &= - \frac{\hat{\theta}}{2} \hat{\pi} , 
\nonumber\\
   \hat{\pi}^{\rho\langle\eta}\hat{\pi}^{\eta\rangle\gamma}\hat{\sigma}_{\rho\gamma}  
   &= - \frac{\hat{\theta}}{4} \hat{\pi}^2 , &
   \hat{\nabla}^{\langle\eta}\hat{\nabla}_{\gamma}\hat{\pi}^{\eta\rangle\gamma} &= 
   \frac{\hat{\theta}^2}{6} \hat{\pi}, 
\nonumber\\ 
   \hat{\nabla}_{\gamma}\hat{\nabla}^{\langle\eta}\hat{\pi}^{\eta\rangle\gamma} 
   &= - \frac{\hat{\theta}^2}{4}\hat{\pi},  &
   \hat{\nabla}^2\hat{\pi}^{\langle\eta\eta\rangle} &= \frac{\hat{\theta}^2}{6} \hat{\pi}.
\end{align} 
Here $\hat{\theta} \equiv 2\tanh\rho$ is the local scalar expansion rate for Gubser flow.

Using the above results the evolution equations for $\hat{\epsilon}$ and $\hat{\pi}$ take the form
\begin{align}
\label{GED}
   \frac{d\hat{\epsilon}}{d\rho} =& - \left( \frac{8}{3} \hat{\epsilon}  - \hat{\pi} \right) \tanh\rho,
\\
\label{Gshear}
   \frac{d\hat{\pi}}{d\rho} =& - \frac{\hat{\pi}}{\hat{\tau}_\pi} 
   + \tanh\rho\left(\frac{4}{3} \hat{\beta}_\pi - \hat{\lambda} \hat{\pi} - \hat{\chi}\frac{\hat{\pi}^2}{\hat{\beta}_\pi}\right). 
\end{align} 
As in the Bjorken case some third-order contributions were brought into the form $\hat{\pi}^2 \hat{\theta}$ by using the first order (Navier-Stokes) relation $\hat{\pi} = (4/3) \hat{\tau}_\pi \hat{\beta}_\pi \tanh\rho$. The transport coefficients in Eq.~(\ref{Gshear}) are given by
\begin{align}
\label{coeffG}
   \hat{\beta}_\pi = \frac{4\hat{P}}{5}, \quad \hat{\lambda} = \frac{46}{21}, \quad \hat{\chi} = \frac{72}{245}. 
\end{align}

For Gubser flow the expression for the LRF entropy density $\hat{s}(\rho)$ is given in terms of $\hat{\pi}$ as
\begin{align}
  \hat{s} = \hat{s}_\mathrm{eq} - \frac{3\,\hat{\beta}}{8\,\hat{\beta}_\pi} \hat{\pi}^2
  + \frac{15\,\hat{\beta}}{168\,\hat{\beta}^2_\pi} \hat{\pi}^3.
\end{align}
Similar to the Bjorken result in Milne coordinates, we find that for Gubser flow the entropy flux vanishes in de Sitter (i.e. LRF) coordinates. 

\section{Anisotropic hydrodynamics}
\label{sec5}

Anisotropic hydrodynamics makes the single particle phase space distribution function $f(x,p)$ explicitly dependent on a spacelike four-vector $l^{\mu}$, which denotes the local anisotropy direction, and a momentum anisotropy parameter $\beta_l$ which controls the amount of deformation from the usual isotropic form. The leading-order part of the distribution function $f(x,p)\equiv f_a(x,p)+\delta\tilde{f}(x,p)$ is in general written as  $f_a\bigl(\beta_u (-u \cdot p), \beta_l (l \cdot p) \bigr)$ \cite{Molnar:2016vvu} such that 
\begin{align}
   \mathrm{lim}_{\beta_l \rightarrow 0  } f_a\bigl(\beta_u (-u \cdot p), \beta_l (l \cdot p)\bigr) 
   = \feq\bigl(\beta_u (-u\cdot p)\bigr).
\end{align}
Different from conventional hydrodynamics, the parameter $\beta_l$ can be arbitrarily large, enabling anisotropic hydrodynamics to handle large deviations of the system from local momentum isotropy and equilibrium. 

In this work, the vector $l^{\mu}$ is taken to point in the longitudinal $\eta$ direction in the LRF, i.e. $l^{\mu} = (0,0,0,1)$ in LRF coordinates, and we consider the widely used Romatschke-Strickland (RS) \cite{Romatschke:2003ms} ansatz for the anisotropic distribution function:
\begin{align}\label{RS}
   f_a \equiv \frs = \exp\left[- \brs\sqrt{p^{\mu}p^{\nu}\Omega_{\mu\nu}}\right],
\end{align}
where
\begin{align}
  \Omega_{\mu\nu}(x) = u^{\mu}(x)u^{\nu}(x) + \xi(x)\, l^{\mu}(x)l^{\mu}(x).
\end{align}
Note that for this choice $\beta_{u}{\,\equiv\,}\brs$ and $\beta_l{\,=\,}\brs\sqrt{\xi}$. The parameter $\brs$ is related to the inverse temperature $\beta = 1/T$ through the Landau matching condition, as we shall see later.

Owing to the presence of an intrinsic directionality $l^{\mu}$ in the system, the energy-momentum tensor $T_\mathrm{RS}^{\mu\nu}$ corresponding to the leading-order distribution $f_a$ only has the general decomposition in the Landau frame \cite{Molnar:2016vvu}
\begin{equation}\label{Ta}
  T^{\mu\nu}_\mathrm{RS} = \epsilon_\mathrm{RS} \, u^{\mu}u^{\nu} + P_{L,\mathrm{RS}} \, l^{\mu}l^{\nu} 
  - P_{T,\mathrm{RS}} \, \Xi^{\mu\nu}.
\end{equation}
Here $\Xi^{\mu\nu} \equiv g^{\mu\nu} + u^{\mu}u^{\nu} - l^{\mu}l^{\nu}$ projects onto the space orthogonal to both $u^{\mu}$ and $l^{\mu}$. The local energy density $\epsilon_\mathrm{RS}$, longitudinal pressure $P_{L,\mathrm{RS}}$, and transverse pressure $P_{T,\mathrm{RS}}$ can be expressed as moments of $\frs$ \cite{Martinez:2017ibh}:
 \begin{align}    
 \label{eRS}
   \epsilon_\mathrm{RS} &= \langle (-u \cdot p)^2  \rangle_\mathrm{RS} = \epsilon(\brs)\, R_{200}(\xi),
\\
\label{PLRS}
   P_{L,\mathrm{RS}} &= \langle (l \cdot p)^2  \rangle_\mathrm{RS} = \epsilon(\brs)\, R_{220}(\xi),
\\
\label{PTRS}
   P_{T,\mathrm{RS}} &= \frac{1}{2}\langle \Xi^{\mu\nu} p_{\mu} p_{\nu} \rangle_\mathrm{RS}
   = \frac{1}{2} P(\brs)\, R_{201}(\xi). 
 \end{align}
 For massless systems they are related by conformal invariance, $\epsilon_\mathrm{RS} = (2P_{T,\mathrm{RS}} + P_{L,\mathrm{RS}})$, and one has $\epsilon(\brs) = 3 P(\brs) = 3/(\pi^2\brs^4)$. The Landau matching condition $\epsilon_\mathrm{RS}(\brs){\,=\,}\epsilon(\beta)$ yields $\beta{\,=\,}\brs/R_{200}^{1/4}$. For a massless Boltzmann gas the anisotropic integrals $R_{nrq}(\xi)$ in Eq.~(\ref{eRS}) can be calculated analytically  \cite{Molnar:2016gwq,Martinez:2017ibh}:
\begin{align}
  R_{200}(\xi) &= \frac{1}{2}\left( \frac{1}{1+\xi} + \frac{\tan^{-1}\sqrt{\xi}}{\sqrt{\xi}} \right)
  \\
  R_{201}(\xi) &= \frac{3}{2\xi} \left( \frac{1}{1+\xi} - (1-\xi)R_{200}(\xi) \right) 
  \\
  R_{220}(\xi) & = -\frac{1}{\xi} \left( \frac{1}{1+\xi} - R_{200}(\xi) \right).
 \end{align}

The residual deviation $\delta\tilde f$ of the distribution function generates, in principle, additional contributions to the longitudinal and transverse pressures, $\delta P_L = \langle (-u \cdot p)^2  \rangle_{\delta\tilde f}$ and $\delta P_T = \langle (l \cdot p)^2  \rangle_{\delta\tilde f}$. We here use the $P_L$ matching scheme \cite{Molnar:2016gwq, Martinez:2017ibh} in which the anisotropy parameter $\xi(x)$ is chosen such that these contributions vanish exactly. This is a dynamical matching scheme similar to Landau matching which defines the local temperature $T(x)$ in such a way that the deviation $\delta f$ from local equilibrium makes no contribution to the energy density $\epsilon(x)$. With this matching scheme we can drop the subscripts RS on $P_L$ and $P_T$.

For massless systems with Bjorken or Gubser symmetry it can be shown that there are no other dissipative contributions from $\delta\tilde f$ to the energy momentum tensor \cite{Molnar:2016gwq, Martinez:2017ibh}. The bulk viscous pressure vanishes by conformal symmetry, and the shear stress tensor is fully specified by the difference between the longitudinal and transverse pressures,
\begin{align}
 \pi^{\mu\nu} = \frac{2(P_L - P_T)}{3} \left( l^{\mu}l^{\nu} - \frac{1}{2}\Xi^{\mu\nu} \right).
\end{align}
It can thus be reduced to a single independent component for which we choose $\pi \equiv -\pi^\eta_\eta = -\tau^2\pi^{\eta\eta} = \frac{2}{3}(P_L{-}P_T)$ in the Bjorken case and $\hat \pi \equiv \hat \pi^{\eta\eta} = \frac{2}{3}(\hat{P}_L{-}\hat{P}_T)$ in the Gubser case. Evolution equations for $\pi$ and $\hat\pi$ are obtained from the Boltzmann equation following Refs.~\cite{Molnar:2016gwq, Martinez:2017ibh}.

For Bjorken flow one finds \cite{Molnar:2016gwq}  that Eqs.~(\ref{BED},\ref{Bshear}) are in anisotropic hydrodynamics replaced by
 \begin{align}
 \label{AHED}
  \frac{d\, \epsilon}{d\tau}  &= - \frac{1}{\tau}\bigl(\epsilon + P_L \bigr),
 \\
 \label{AHPL}
  \frac{dP_L}{d\tau} &= - \frac{ P_L{-}P}{\tau_\pi} + \frac{1}{\tau} \Bigl( 3 P_L - I^\mathrm{RS}_{240} \Bigr).
 \end{align}
Here we followed \cite{Molnar:2016gwq} and expressed the shear stress $\pi$ through the longitudinal pressure $P_L$ via $\pi=P{-}P_L=\frac{1}{3}\epsilon{-}P_L$. The thermodynamic integral $I_{240}^\mathrm{RS}$ over the RS distribution function is given in terms of the momentum deformation parameter $\xi$ as $I^\mathrm{RS}_{240}(\beta,\xi) =  \epsilon(\beta) \, R_{240}(\xi)/R_{200}(\xi)$, with
 \begin{align}
   R_{240}(\xi) =  \frac{1}{\xi^2} \left( \frac{3+\xi}{1+\xi}  - 3 R_{200}(\xi) \right).
 \end{align}
Eq.~(\ref{AHED}) agrees with Eq.~(\ref{BED}) in Sec.~\ref{sec4b} while the evolution equations for the shear stress $\pi=P{-}P_L$, Eqs.~(\ref{Bshear}) and (\ref{AHPL}), differ. We solve Eqs.~(\ref{AHED},\ref{AHPL}) by using the relations (\ref{eRS},\ref{PLRS}) to write $P_L = \epsilon(\beta) R_{220}(\xi)/R_{200}(\xi)$ and convert Eq.~(\ref{AHPL}) into an evolution equation for $\xi$.

For Gubser flow one obtains \cite{Martinez:2017ibh} the energy conservation law (\ref{GED}) and, instead of Eq.~(\ref{Gshear}), the shear stress evolution 
\begin{align}
\label{AHshear}
 \frac{d \hat{\pi}}{d\rho} = -  \frac{\hat{\pi}}{\hat{\tau}_\pi} +
 \tanh\rho\left( \frac{4}{3} \hat{\beta}_\pi -\hat\lambda_a\hat\pi
 -\hat{I}_{240} \right),
\end{align}
with $\hat{I}_{240}(\hat\beta,\xi) =  \hat\epsilon(\hat\beta) \, R_{240}(\xi)/R_{200}(\xi)$ and the modified transport coefficient $\hat\lambda_a=\frac{4}{3}$.

For the definition of the out-of-equilibrium entropy current we substitute $f(x,p) = \frs(x,p)+\delta\tilde f$ in Eq.~(\ref{EFC}):
\begin{align}
 s = \int dp \, (u \cdot p) f (\ln f -1 ) = s_a +\delta\tilde s.
\end{align}
Using $\frs=\exp(-\brs\sqrt{(1{+}\xi)w^2 /\tau^2 + p_T^2})$ for Bjorken flow, together with the integration measure $dp = dw\, d^2 p_T/[(2\pi)^3 \tau (-u{\cdot}p)]$, and applying the transformation $w\to w' = w\sqrt{1+\xi}$ for which $\frs\to\feq(\tau,p_T,w';\brs)$, the leading contribution $s_a$ can be evaluated exactly:
\begin{align}
\label{RSentropy}
  s_a(\tau) 
  &= -\frac{1}{\tau} \int \frac{d w \, d^2 p_T}{(2\pi)^3} \frs \bigl( \ln\frs -1 \bigr)
\\
  & = \frac{4}{\pi^2 \beta^3_\mathrm{RS}}\frac{1}{\sqrt{1+\xi}}.
\end{align}
To linear order the $\delta\tilde f$ correction to the entropy density is given by
\begin{align}
\label{RSentropy}
  \delta\tilde s &=  \int dp \, (u \cdot p) \, \delta\tilde f \,\ln\frs  + \mathcal{O}\bigl((\delta\tilde f)^2\bigr)
\\ 
  &\approx -\brs \int dp\, (u\cdot p)\, \sqrt{(u\cdot p)^2+\xi\,(l\cdot p)^2}\ \delta\tilde f.
\end{align}
To evaluate it an approximate solution of the Boltzmann equation for $\delta\tilde f$ is needed. We here use the moments method \cite{Denicol:2012cn} in the 14-moment approximation, $\delta\tilde f\approx\delta\tilde f_{14}$. Due to our matching conditions, for Bjorken and Gubser flows $\delta\tilde f$ contributes zero to all 14 hydrodynamic moments of the distribution function, hence $\delta\tilde f_{14} {\,=\,}0$.\footnote{%
    The general form of the 14-moment approximation $\delta\tilde f_{14}$ for systems with Gubser symmetry
    is given in Eq.~(48) of Ref.~\cite{Martinez:2017ibh}. For systems with Bjorken symmetry the same expression
    holds without the hats. It is straightforward to see that in the $P_L$-matching scheme this expression is zero
    in both cases.
    }
%
\begin{figure*}[!htb]
 \begin{center}
    \includegraphics[width=0.95\linewidth]{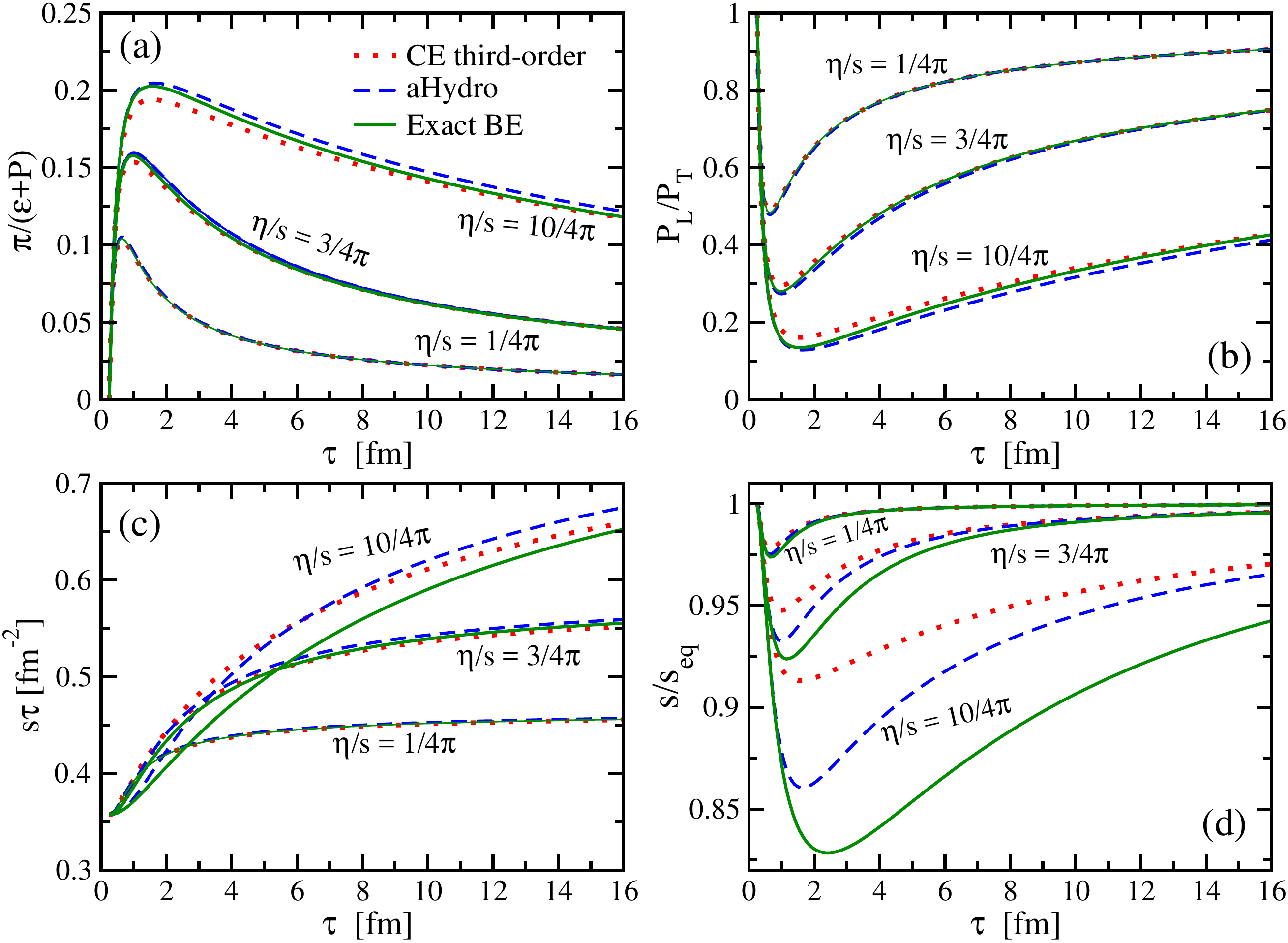}
 \end{center}
 \vspace{-0.5cm}
 \caption{(Color Online). Proper time evolution of (a) the normalised shear stress $\pi/(\epsilon{+}P)$, 
               (b) the pressure anisotropy $P_L/P_T$, (c) the entropy density per unit rapidity and transverse 
               area $s\tau$, and (d) the normalised entropy density $s/s_\mathrm{eq}$, for Chapman-Enskog 
               third-order hydrodynamics (dotted red lines), anisotropic hydrodynamics (dashed black lines) 
               and for the exact solution of the RTA Boltzmann equation (solid green lines). For each theory three 
               sets of curves are shown, corresponding to three different values for the specific shear viscosity,
               $4\pi\eta/s{\,=\,}1,\,3,$ and 10. All curves assume an initial equilibrium state (i.e. $\pi_0 = 0$) 
               with temperature $T_0 = 300$\,MeV at $\tau_0 = 0.25$\,fm/$c$.}
 \label{F1}
\end{figure*}
%
This shows that in the $P_L$-matching scheme only non-hydrodynamic moments of the distribution function contribute to the residual non-equilibrium entropy density $\delta\tilde s$. We leave a detailed study of such non-hydrodynamic mode contributions to entropy production to future work.

For Gubser flow a similar calculation yields for the leading contribution
\begin{align}
   \hat{s}_a(\rho) = \frac{4}{\pi^2 \hat{\beta}_\mathrm{RS}^3} \frac{1}{\sqrt{1+\xi}}.
\end{align}
The correction $\delta\hat{\tilde s}$, at linear order in $\delta\tilde f$, again vanishes in the 14-moment approximation.

\section{Numerical results and discussion}
\label{sec6}

We compare the numerical results obtained from three different formalisms: Chapman-Enskog third-order viscous hydrodynamics, anisotropic hydrodynamics with $P_L$ matching, and the exact solution of the RTA Boltzmann equation. Although in principle each set of evolution equations can be solved for any initial condition, we here only show results evolving from local thermal equilibrium (with vanishing initial momentum-space deformation $\xi_0=0$ and shear stress $\pi_0=0$) at some initial time.

\subsection{Bjorken flow}
\label{sec6a}

\begin{figure*}[!htb]
 \begin{center}
    \includegraphics[width=0.95\linewidth]{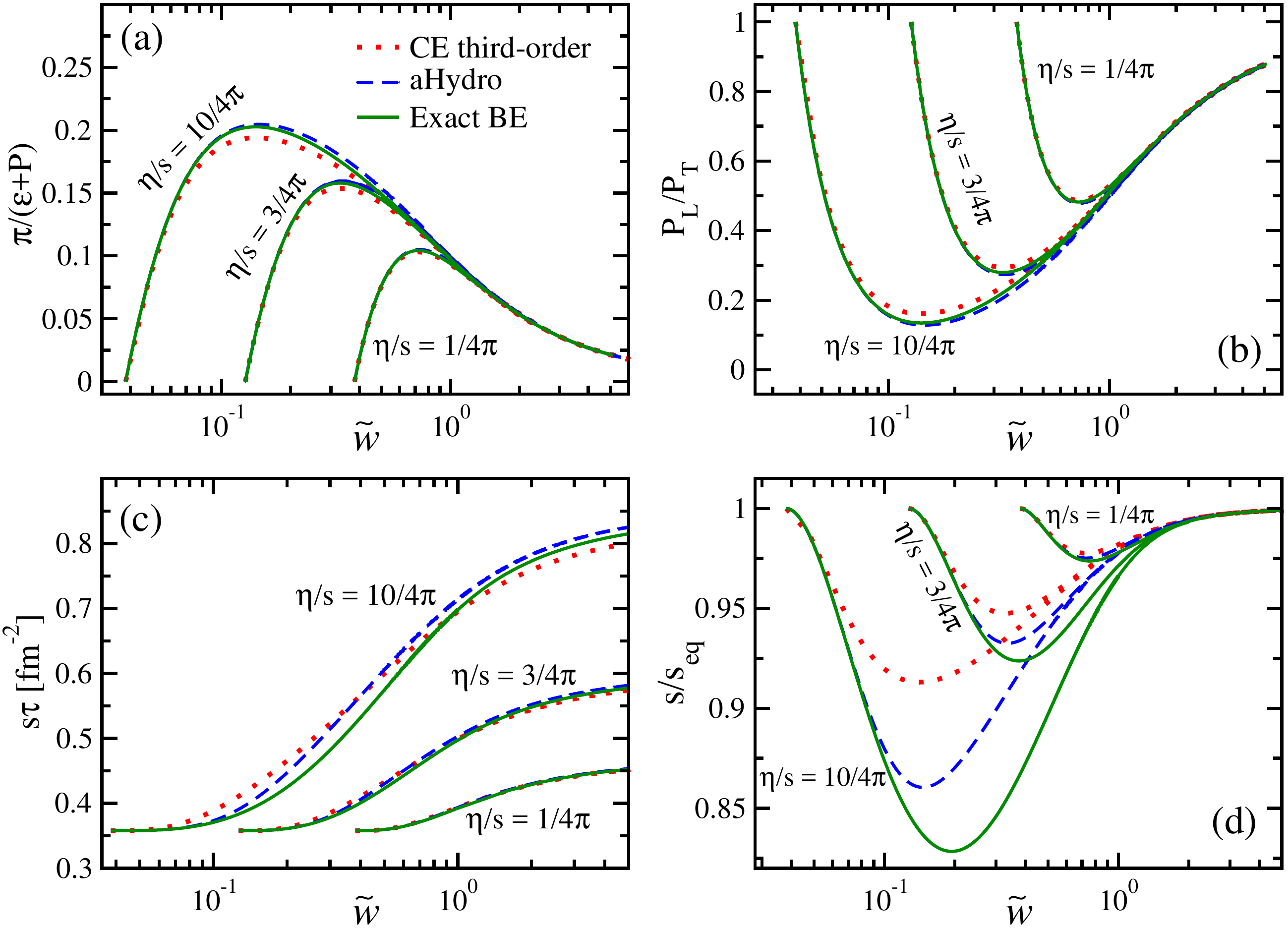}
 \end{center}
 \vspace{-0.5cm}
 \caption{(Color Online). The same quantities as shown in Fig.~\ref{F1}, but now plotted as a function of
 	the scaling variable $\tilde{w}\equiv\tau T/(4\pi\eta/s)$. See text for discussion.}
 \label{F2}
\end{figure*}

For our Bjorken flow results we initialize the system at longitudinal proper time $\tau_0=0.25$\,fm/$c$ with initial temperature $T_0 = 300$\,MeV. Figure~\ref{F1} shows the resulting proper time evolution of the normalised shear stress $\pi/(\epsilon+P)$ (panel a), the pressure anisotropy $P_L/P_T \equiv (P-\pi)/(P{+}\pi/2)$ (panel b), the entropy density per unit rapidity and transverse area $s\tau$ (panel c), and the normalised entropy density $s/s_\mathrm{eq}$ (panel d), for the three theories listed above and three choices of the specific shear viscosity as indicated in the figure.

For small specific viscosity $\bar{\eta}\equiv\eta/s = 1/4\pi$, all three formalisms yield very similar results. Except for the normalized entropy density $s/s_\mathrm{eq}$ in Fig.~\ref{F1}d, the three curves agree within line thickness. As the specific shear viscosity increases, increasing differences between the three formalisms become visible. Generally the differences remain small for the hydrodynamic moments of the distribution function, i.e. for the evolution of the normalized shear stress $\bar{\pi}\equiv\pi/(\epsilon{+}P)$ and pressure anisotropy (which, because of $P_L/P_T= (1{-}4\bar\pi)/(1+2\bar\pi)$, are basically the same quantity). Third-order Chapman-Enskog hydrodynamics performs somewhat better at late times whereas anisotropic hydrodynamics reproduces the exact solution more accurately at earlier times; the exact solution lies between these two hydrodynamic approximations. 

As seen in Figs.~\ref{F1}c and \ref{F1}d, the differences between macroscopic hydrodynamic and exact microscopic kinetic evolution are larger for the entropy. In ideal fluid dynamics with Bjorken flow, $s\tau$ is a constant of motion. The increase of $s\tau$ with time shown in Fig.~\ref{F1}c thus illustrates the rate of entropy production by dissipative effects in the different approaches. One sees that non-equilibrium effects on the rate of entropy production are not as well described by the hydrodynamic models as is the non-equilibrium evolution of the energy-momentum tensor shown in panels a and b. Fig.~\ref{F1}d shows the non-equilibrium deviation of the entropy from the value expected from the first law of thermodynamics, $s_\mathrm{eq}=(\epsilon{+}P)/T=4P/T$ (where both $P$ and $T$ evolve according to viscous fluid dynamics). For $\eta/s = 10$ times the ``minimal'' KSS value of $1/(4\pi)$ \cite{Policastro:2001yc}, the entropy differs from the ``equilibrium'' value $s_\mathrm{eq}$ by up to 10\% for third-order Chapman-Enskog hydrodynamics, and even for anisotropic hydrodynamics (where, as discussed at the end of the previous section, only non-hydrodynamic moments of the distribution function contribute to the residual entropy) the deviation is still 5-7\% over most of the evolution history. This indicates that, while the coupling of non-hydrodynamic modes into the evolution of the hydrodynamic moments of the distribution function is rather weak, the same is not true for the entropy density.   

A remarkable feature of the entropy evolution predicted by the exact solution of the RTA Boltzmann equation is the crossing of the three green curves in Fig.~\ref{F1}c corresponding to different values of $\bar{\eta}$: As the value of $\bar\eta$ increases, the initial rate of entropy production decreases, but entropy is produced over a longer time period such that its eventual saturation value increases with $\bar\eta$. This feature, which is shared by aHydro, but not by the third-order Chapman-Enskog approach, appears counter-intuitive at first sight: In first-order Navier-Stokes theory, the slope of $s\tau$ as a function of $\tau$ is proportional to $\bar{\eta}$: $d(s\tau)/d\tau = 4\bar{\eta}s/3\tau T$. However, this argument implicitly assumes the equilibrium definition of the entropy density, $s\equiv s_\mathrm{eq} = (\epsilon+P)/T$, and the substantial deviation of the exact result from this lowest-order expectation illustrated in Fig.~\ref{F1}d (which shows that the deviation {\em increases} with increasing $\bar\eta$) demonstrates the importance of higher order terms in the definition of the entropy density. (Note that both aHydro and the third-order Chapman-Enskog approach have trouble accounting for this non-equilibrium deviation of the entropy from the first law of thermodynamics.) Microscopically, the increasing deviation from naive Navier-Stokes expecations is related to the growth of the relaxation time with increasing $\bar{\eta}$, resulting in a slower response to the expansion driving the system away from equilbrium. We have checked that the curve crossing disappears when plotting $s_\mathrm{eq} \tau$ instead of $s \tau$; in this case the initial slope of the curves is directly proportional to  $\bar{\eta}$. 

Larger values of the specific shear viscosity $\bar\eta$ lead to stronger viscous heating, thereby delaying the cooling by expansion of the fireball. At a given (sufficiently late) proper time $\tau$ the more viscous fluid thus has a higher temperature than the less viscous one if both started out with the same initial temperature $T_0$. The authors of Ref.~\cite{Heller:2016rtz} showed that this effect can be scaled out of the evolution plots for dimensionless ratios such as $\pi/P$ or $P_L/P_T$ if one plots them as a function of the dimensionless scaling variable $\tilde{w}=\tau T/(4\pi\eta/s)$ instead of $\tau$. Fig.~\ref{F2} shows this for the four quantities plotted in Fig.~\ref{F1}. The dimensionless ratios $\pi/(\epsilon{+}P)$, $P_L/P_T$, and even the non-equilibrium entropy ratio $s/s_\mathrm{eq}$ exhibit clear scaling behavior, converging at around $\tilde{w}\simeq 1$ to a universal late-time attractor given by relativistic Navier-Stokes theory. That the aHydro attractor, whose equation involves a resummation of terms in powers of inverse Reynolds number, closely matches with the exact attractor has already been demonstrated in Ref.~\cite{Strickland:2017kux}, albeit with a slightly different version of aHydro that did not implement $P_L$-matching. We note that the dimensionful quantity $s\tau$ does not scale, but the crossing of the curves seen in Fig.~\ref{F1}c is removed by rescaling the time evolution variable. The scaling plots shown in Fig.~\ref{F2} reinforce the observation made in Fig.~\ref{F1} that the hydrodynamic approximations reproduce the exact evolution of the energy momentum tensor, in particular the normalized shear stress and pressure anisotropy, much more accurately than that of the entropy ratio $s/s_\mathrm{eq}$. Eventually, however, even this latter ratio approaches a universal Navier-Stokes attractor, albeit only at $\tilde{w}\gtrsim 2$, i.e. twice later than the hydrodynamic moments.       

\begin{figure}[b]
 \begin{center}
    \includegraphics[width=\linewidth]{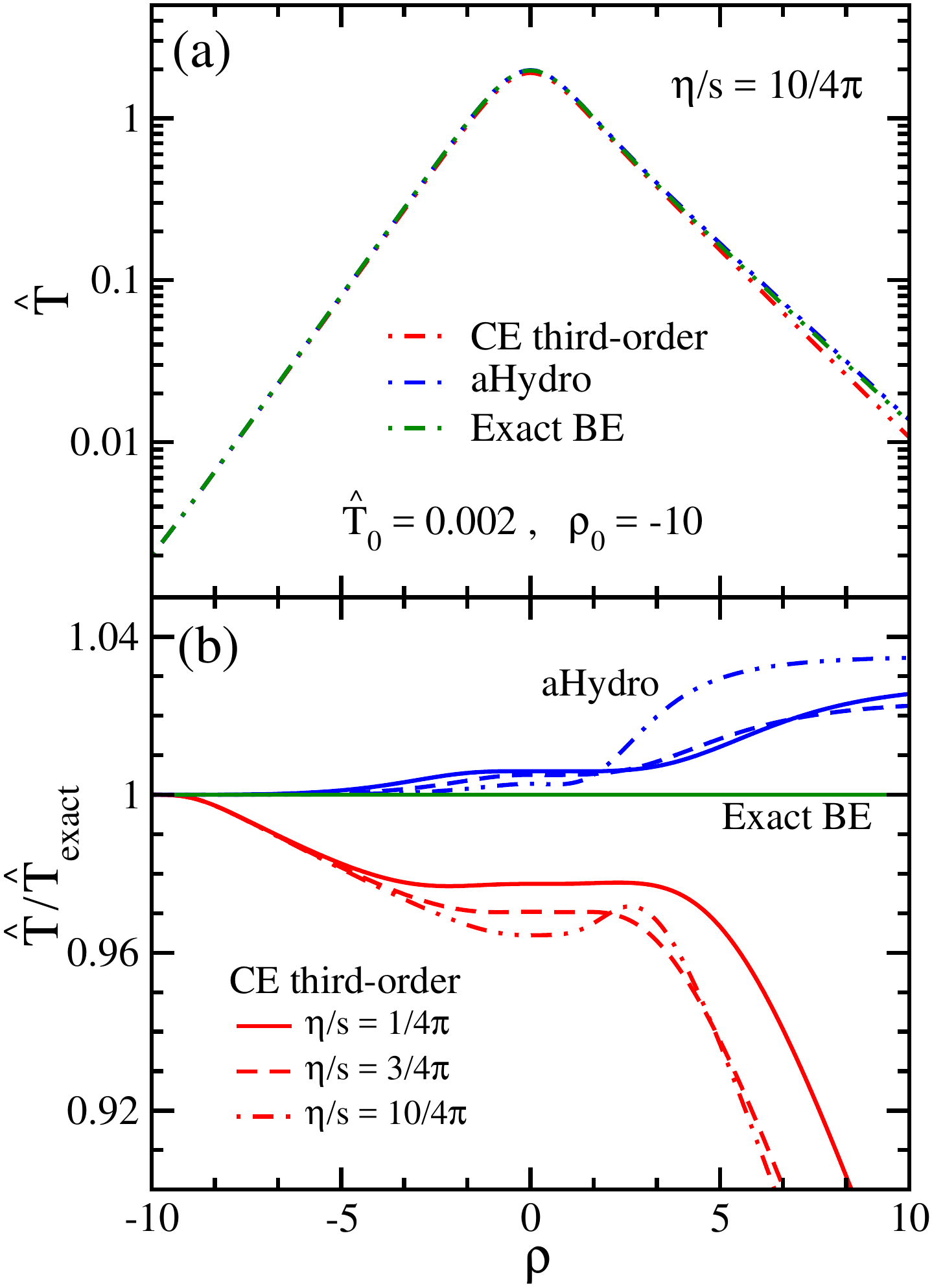}
 \end{center}
 \vspace{-0.5cm}
 \caption{(Color Online). de Sitter time evolution the (normalized) temperature $\hat{T}$ for Gubser flow,
 	in absolute terms (a) and relative to the temperature corresponding to the energy density associated
	with the exact solution of the RTA Boltzmann equation (b). The results for anisotropic hydrodynamics
	(blue) and the third-order Chapman-Enskog approach (red) are compared with the exact solution (green).
	Panel (a) shows results for $\eta/s=10/(4\pi)$ only whereas in panel (b) results are compared for three 
	different values of the specific shear viscosity, $4\pi\eta/s=1$, 3, and 10.}
 \label{F3}
\end{figure}

\subsection{Gubser flow}
\label{sec6b}

\begin{figure*}[!bht]
 \begin{center}
    \includegraphics[width=0.95\linewidth]{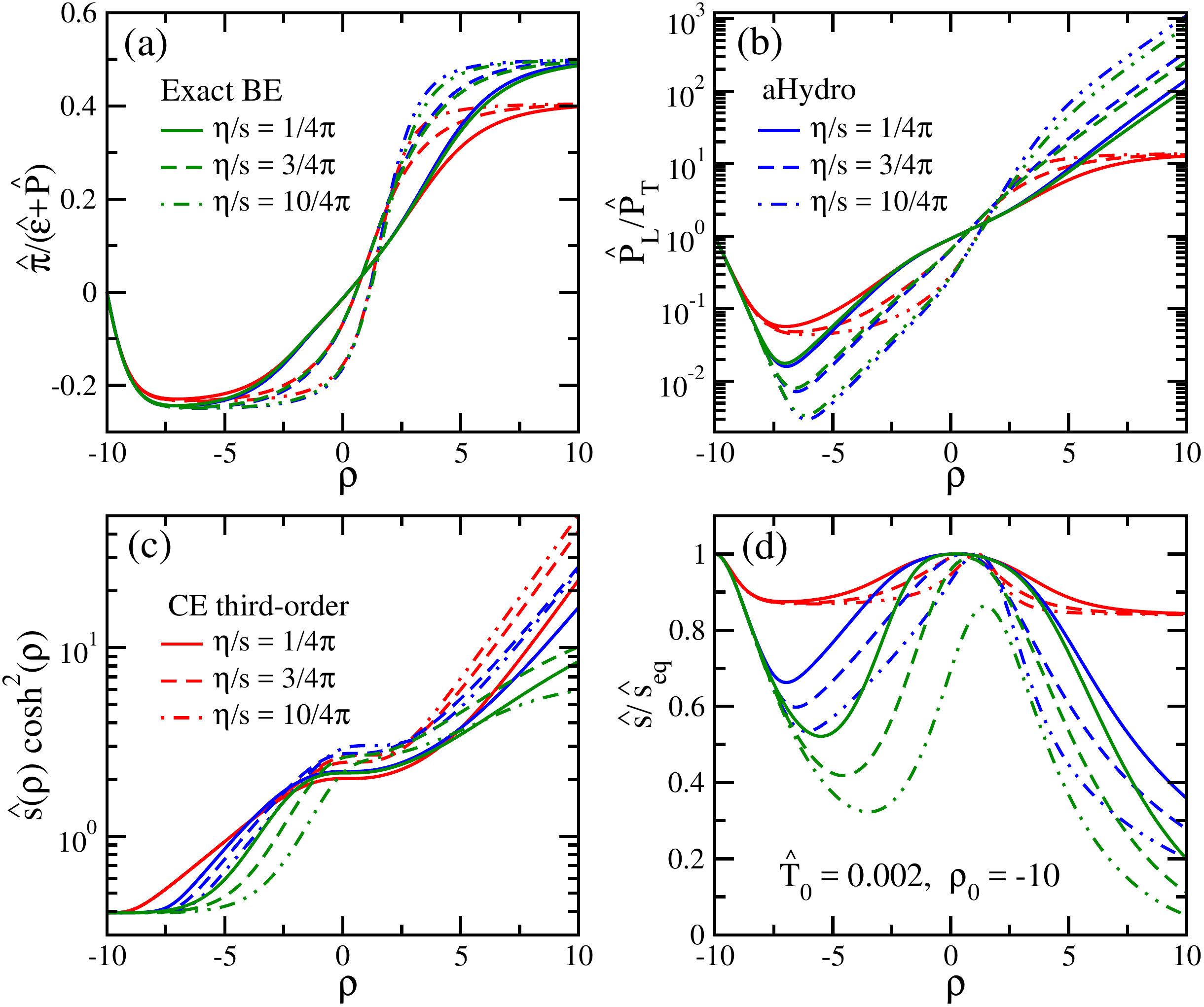}
 \end{center}
 \vspace{-0.5cm}
 \caption{(Color Online). de Sitter time evolution of (a) the normalised shear stress 
            $\hat{\pi}/(\hat{\epsilon}{+}\hat{P})$, (b) the pressure anisotropy $\hat{P}_L/\hat{P}_T$, 
            (c) the entropy content $\hat{s}\cosh^2(\rho)$, and (d) the normalised entropy density 
            $\hat{s}/\hat{s}_\mathrm{eq}$, for Chapman-Enskog third-order hydrodynamics (red), 
            anisotropic hydrodynamics (blue), and for the exact solution of the RTA Boltzmann equation (green).
            For each theory three sets of curves are shown, corresponding to three different values for the
            specific shear viscosity, $4\pi\eta/s{\,=\,}1$ (solid), 3 (dashed), and 10 (dash-dotted).
            Thermal equilibrium initial conditions ($\hat\pi=0$) with initial temperature $\hat{T}_0 = 0.002$ 
            were implemented at $\rho_0 = -10$.}
 \label{F4}
\end{figure*}

Gubser flow is interesting because of its very strong transverse expansion which asymptotically (i.e. for very large de Sitter times) drives the system arbitrarily far away from local thermal equilibrium, into a state of free-streaming
\cite{Denicol:2014xca,Denicol:2014tha}. This is in contrast to Bjorken flow where there is no transverse flow and the longitudinal expansion rate decreases for late longitudinal proper times, allowing the system to settle into a state of approximate local thermal equilibrium. The dramatic transverse expansion encoded in Gubser flow thus provides a testbed for the performance of hydrodynamic approximations in situations very far from equilibrium.   

For Gubser flow we initialize the system in equilibrium (i.e. with $\hat{\pi}_0{\,=\,}\xi_0{\,=\,}0$) at de Sitter time $\rho_0{\,=\,}{-}10$ with initial normalized temperature $\hat{T}=0.002$.\footnote{%
	For a typical transverse size of $1/q = 4.3$\,fm, this corresponds to an initial temperature 
	$T \approx 2$\,GeV at $\tau \approx 1.95\times10^{-4}$\,fm in the fireball center.
	} 
%
\begin{figure*}[!htb]
 \begin{center}
    \includegraphics[width=0.95\linewidth]{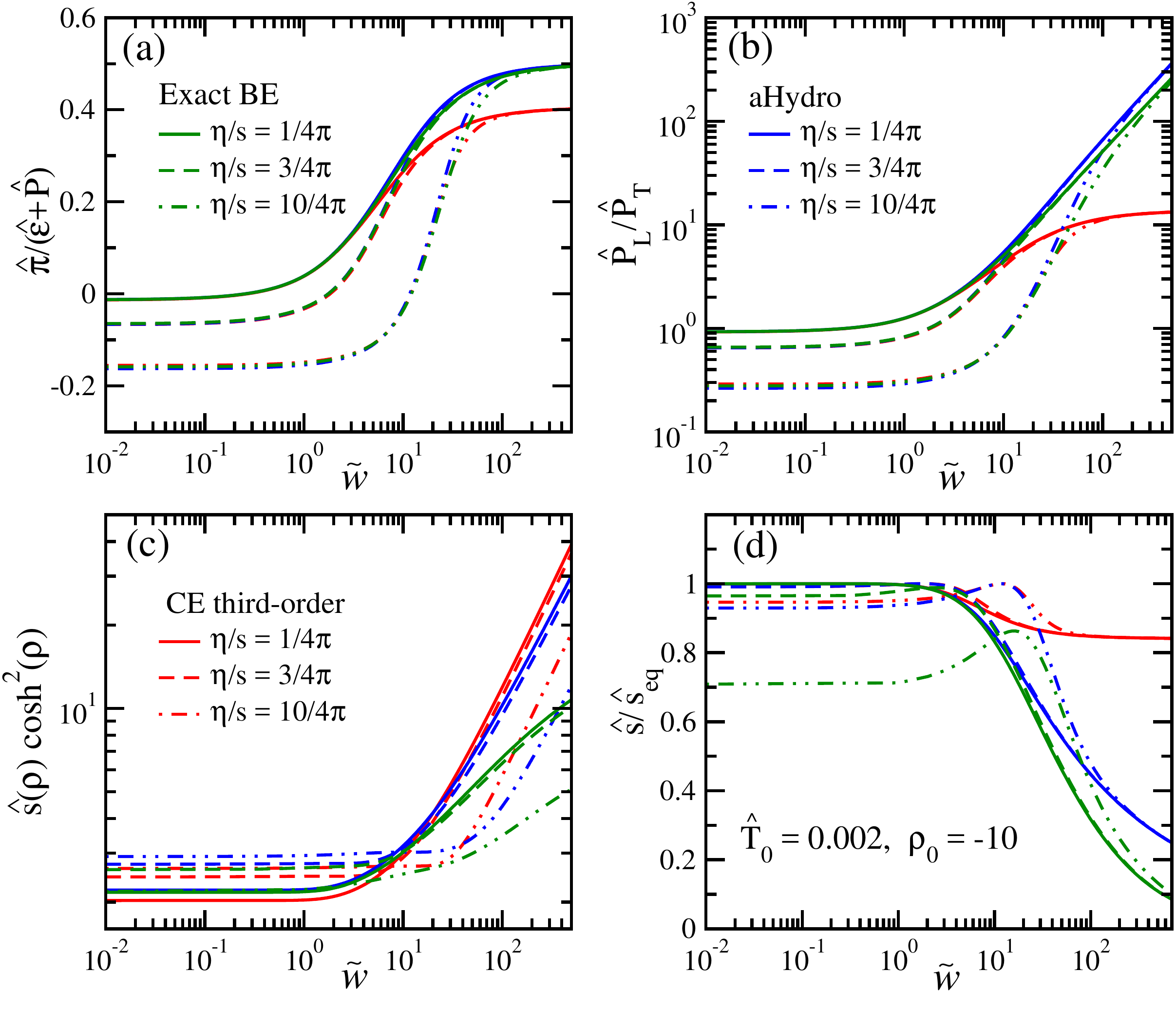}
 \end{center}
 \vspace{-0.5cm}
 \caption{(Color Online). The same quantities as shown in Fig.~\ref{F4}, but now plotted as a function of
 	the scaling variable $\tilde{w}\equiv(4\pi\eta/s)(2\tanh\rho)/\hat{T}$. See text for discussion.}
 \label{F5}
\end{figure*}
%
The temperature evolution is shown in Fig.~\ref{F3}.\footnote{%
	Astute readers may notice a slight discrepancy between the Gubser flow curves shown in Fig.~\ref{F3}b
	and the corresponding curves for aHydro with $P_L$-matching shown in Fig.~1 of Ref.~\cite{Martinez:2017ibh}.
	This difference is of numerical origin: In Ref.~\cite{Martinez:2017ibh} the reference curves for the exact
	solution of the RTA Boltzmann equation were computed with not quite sufficient numerical resolution, 
	resulting in discrepancies for the temperature $\hat{T}$ of up to 1.2\% from the fully converged results shown 
	here.
	}
As discussed in \cite{Martinez:2017ibh}, at early de Sitter times the system rapidly moves away from the initial equilibrium state as a result of rapid initial longitudinal expansion (resulting in negative pressure anisotropy $\hat{\pi}\sim\hat{P}_L{-}\hat{P}_T<0$), then briefly passes through a transient state of approximate local momentum isotropy ($\hat\pi=\xi=0$) before again being driven away from it by increasingly strong transverse expansion (resulting in positive pressure anisotropy $\hat{\pi}\sim\hat{P}_L{-}\hat{P}_T>0$), eventually leading to free-streaming with $\hat{\pi}/(\hat\epsilon{+}\hat{P}) \to 0.5$ at late de Sitter times. Fig.~\ref{F3} shows that at late de Sitter times anisotropic hydrodynamics slightly overpredicts the temperature corresponding (by Landau matching) to the energy density of the exact solution of the RTA Boltzmann equation, by a constant factor. For the third-order Chapman-Enskog approach, the asymptotic temperature is seen to keep falling further and further below that of the exact solution, indicating (as the following figures will show more clearly) that this hydrodynamic model {\em does not} correctly approach the asymptotic free-streaming state and underpredicts the shear stress and viscous heating at late de Sitter times. For both hydrodynamic approximations the asymptotic deviation from the exact solution increases with the specific shear viscosity $\bar\eta$. 

In Fig.~\ref{F4} we show the de Sitter time evolution of the Gubser analogues of the quantities plotted in Fig.~\ref{F1} above for Bjorken flow. As already reported in \cite{Martinez:2017ibh}, anisotropic hydrodynamics with $P_L$ matching provides a very accurate approximation to the exact solution of the RTA Boltzmann equation for the evolution of the shear stress and pressure anisotropy (panels a and b). In particular, it approaches the correct free-streaming limit at large de Sitter times. This approach is faster (in $\rho$) for larger specific shear viscosity $\eta/s$. However, as was the case for Bjorken flow, the ability of aHydro to describe the evolution of the entropy content of the system (panel c) and of the non-equilibrium correction to the first law of thermodynamics (shown in panel d) is much more limited. Especially at late de Sitter times, the aHydro curves appear to move farther and farther away from the exact solution. 

For the third-order Chapman-Enskog approach, large deviations from the exact solution at late de Sitter times are even observed for the hydrodynamic moments shown in Figs.~\ref{F4}a and b: Instead of saturating 
at the free-streaming limit $\hat{\bar\pi}\equiv\hat{\pi}/(\hat\epsilon{+}\hat{P}) = 0.5$, the normalized shear stress in Fig.~\ref{F4}a saturates at 0.4. As a result, the pressure anisotropy $\hat{P}_L/\hat{P}_T=(1{+}4\hat{\bar\pi})/(1{-}2\hat{\bar\pi})$ shown in Fig.~\ref{F4}b saturates at large de Sitter times in the third-order Chapman-Enskog approach instead of continuing to grow as dictated by the exact solution of the RTA Boltzmann equation and is correctly reproduced by aHydro with $P_L$-matching. This failure is similar to the one observed in DNMR theory \cite{Denicol:2012cn} (which is a second-order viscous hydrodynamic approach based on an expansion around a locally isotropic momentum distribution function) except that in DNMR theory $\hat{\bar\pi}$ saturates at a value $>0.5$, corresponding to negative transverse pressure and instability against cavitation \cite{Martinez:2017ibh}.

As far as the de Sitter time evolution of the entropy content of the system (Fig.~\ref{F4}c) and of the non-equilibrium correction to the first law of thermodynamics (Fig.~\ref{F4}d) are concerned, the discrepancies between third-order Chapman-Enskog hydrodynamics and the exact solution of the Boltzmann equation are even larger than those observed for anisotropic hydrodynamics. Although all approaches correctly predict that the entropy density $\hat{s}\cosh^2\rho$ increases as dictated by the second law of thermodynamics (Fig.~\ref{F4}c), the rate of increase is overpredicted by the hydrodynamic models at late de Sitter times. The rate of viscous entropy production is controlled by the normalized shear stress $\hat{\bar\pi}$ shown in Fig.~\ref{F4}a; near $\rho=0$ it is small in all three approaches because $\hat{\bar\pi}$ passes through zero. Finally, Fig.~\ref{F4}d shows that third-order Chapman-Enskog hydrodynamics predicts a saturation of the ratio $s/s_\mathrm{eq}$ at large de Sitter times whereas the exact solution shows that this ratio should continue to decrease as $\rho$ keeps increasing. aHydro reproduces this continued decrease, but at an incorrect rate. 

It is worth noting that for both hydrodynamic approximations studied here, the ratio $s/s_\mathrm{eq}$ shown in Fig.~\ref{F4}d passes through 1 near $\rho=0$ where $\hat{\bar\pi}$ passes through zero. This is not the case for the exact solution which shows non-vanishing deviations of this ratio from unity (whose magnitude increases with $\eta/s$) even when $\hat{\bar\pi}=0$. As similar observation was made before in Ref.~\cite{Bazow:2016oky}, it shows that the exact solution of the Boltzmann equation includes contributions to the non-equilibrium entropy from non-hydrodynamic moments \cite{Bazow:2016oky} that are not captured by the hydrodynamic approximations studied here.

We close this section by replotting Fig.\,\ref{F4} as a function of the scaling variable $\tilde{w}=(4\pi\eta/s)(2\tanh\rho)/\hat{T}$ \cite{Behtash:2017wqg} in Fig.\,\ref{F5}.\footnote{%
	Noting that $2\tanh\rho$ is the scalar expansion rate of Gubser flow, corresponding to $1/\tau$ in Bjorken
	flow, one sees that this definition of $\tilde{w}$ is the inverse of the definition used for Bjorken flow in the 
	preceding subsection. We have included the factor $4\pi\eta/s$ in the definition of $\tilde{w}$ in order to 
        scale out the $\eta/s$-dependence of viscous heating in the Navier-Stokes limit of small shear stresses 
	\cite{Heller:2016rtz}.
	}
Our findings are consistent with the detailed study of the Gubser flow fixed point presented in Ref.~\cite{Behtash:2017wqg}. As for the case of Bjorken flow, one observes convergence of the curves describing the evolution of the normalized shear stress (Fig.~\ref{F5}a) and pressure anisotropy (Fig.~\ref{F5}b) for different specific shear viscosities to a common attractor at large values of $\tilde{w}$.\footnote{%
	We note that without including the factor $\eta/s$ in the definition of the scaling variable $w$ the 
	pressure anisotropy $P_L/P_T$ approaches different late-time attractors for different values of 
	$\eta/s$ (not shown). This pressure anisotropy diverges at large $\rho$, $w$ and $\tilde{w}$ as the
	system approaches free-streaming and the transverse pressure goes to zero. Including the factor
	$\eta/s$ in the definition of the scaling variable $\tilde{w}$ exhibits an additional degree of universality 
	in this asymptotic behavior that is not seen when plotting the pressure anisotropy as a function of $w$.
	} 
In this case, however, the attractor for the normalized shear stress $\hat{\bar\pi}$ differs for third-order Chapman-Enskog hydrodynamics from the shared ``free-streaming attractor" for aHydro and the exact solution of the RTA Boltzmann equation. This reflects the above observation that third-order Chapman-Enskog hydrodynamics does not approach the correct free-streaming limit at large de Sitter times. Fig.~\ref{F5}b additionally shows that the rate at which the trajectories for aHydro and the RTA Boltzmann equation approach the asymptotic value $\hat{\bar\pi}=0.5$ is slightly different for the two theories, but insensitive to the value of $\eta/s$ in each case.

In contrast to the dimensionless ratios shown in panels a, b, and d, the evolution of the dimensionful entropy density shown in Fig.~\ref{F5}c exhibits no clear scaling behavior. For the non-equilibrium entropy ratio $s/s_\mathrm{eq}$ in Fig.~\ref{F5}d one observes different attractors for all three dynamical approaches: whereas in each case the curves corresponding to different specific shear viscosity converge at large $\tilde{w}$, the attractors they converge to are very different for the exact solution, aHydro and third-order Chapman-Enskog. The difference between the aHydro and exact attractors is smaller than between third-order Chapman-Enskog and the exact result, but still large. Clearly, the hydrodynamic approximations are having difficulties reproducing the non-equilibrium contributions to the entropy density at large $\tilde{w}$, i.e. deep in the free-streaming region of the exact solution. 
	 
\section{Conclusions}
\label{sec7}

In this work, we have considered two different formalisms for deriving macroscopic descriptions of the non-equilibrium dynamics of a system, namely, dissipative hydrodynamics using the Chapman-Enskog iterative scheme to third order and anisotropic hydrodynamics (aHydro) with $P_L$-matching. The performance of these different hydrodynamic schemes was tested by comparing their predictions with the exact solution of the RTA Boltzmann equation in two situations where such an exact solution is available, namely for the Bjorken and Gubser flows. Both situations are effectively one-dimensional such that the energy-momentum tensor can be characterized by just two hydrodynamic moments of the microscopic distribution function, the energy density (or, equivalently, the temperature) and a single shear stress component. The shear stress also defines the phenomenologically important longitudinal-transverse pressure anisotropy $P_L/P_T$. Bjorken and Gubser flows describe two extreme situations that bracket realistic situations: while both share boost-invariant longitudinal expansion, Bjorken flow lacks any transverse expansion (and correspondingly allows the system to approach a state of local thermal equilibrium at late times) whereas Gubser flow features very strong radial expansion in the transverse directions which at late times drives the system completely away from local equilibrium into an asymptotic state of free-streaming. Both flows start out with strong longitudinal expansion in which dissipative effects deform the local rest frame momentum distribution by making it narrower in the longitudinal momentum $p_\eta$ than in transverse momentum $p_T$ (such that $P_L{-}P_T<0$), but for Bjorken flow the local momentum distribution becomes asymptotically isotropic whereas for Gubser flow it eventually becomes narrower in $p_T$ than $p_\eta$ (leading to $P_L{-}P_T>0$). The two flows thus present a testbed for macroscopic hydrodynamic approximations of the microscopic dynamics under very different conditions of anisotropic expansion, with opposite signs of the pressure anisotropy $P_L-P_T$ at late times. 

In addition to the evolution of the abovementioned hydrodynamic moments (whose dynamics has been studied before) we also explored here the evolution of the entropy density of the system (which, in practical situations such as relativistic heavy-ion collisions, controls the multiplicity of finally emitted hadrons). Our interest in the entropy arises from previous observations \cite{Bazow:2016oky} that suggested that the entropy evolution is more strongly influenced by dynamical couplings to non-hydrodynamic moments of the distribution function, and we wanted to know how well these couplings can be captured in macroscopic hydrodynamic treatments. 

In all cases (i.e. for both anisotropic flow patterns and for all the observables studied) we found that anisotropic hydrodynamics with $P_L$-matching provides a more accurate approximation to the exact evolution obtained from the exact solution of the Boltzmann equation than does the dissipative hydrodynamics derived from a third-order Chapman-Enskog expansion of the distribution function. The latter is found to consistently under-predict the deviation from local equilibrium even when terms up to third-order in velocity gradients are kept in the expansion of the shear stress tensor. As a consequence, the results of third-order Chapman-Enskog hydrodynamics deviate substantially from the exact solution whenever momentum-space anisotropies become non-perturbatively large. This feature is most apparent for Gubser flow, both during early times when longitudinal expansion dominates the pressure anisotropy and at late times when the strong transverse expansion drives the pressure anisotropy and pushes the system towards free-streaming.

Our findings can be understood most intuitively when plotting them against a dimensionless time variable $\tilde{w}$ (defined in the text) that is scaled by the microscopic relaxation time (which increases with increasing specific shear viscosity $\eta/s$). For Bjorken flow one finds that the exact solution and the two hydrodynamic approximation schemes studied in this paper share a common attractor to which all solutions converge at late times, irrespective of initial conditions. For Gubser flow, aHydro shares a common attractor with the exact solution for the normalized shear stress and pressure anisotropy, whereas these quantities approach a different attractor for third-order Chapman-Enskog hydrodynamics. For the non-equilibrium entropy, the asymptotic evolution in Gubser flow is controlled by three different attractors for the exact solution and the two hydrodynamic approximation schemes, with the differences between the exact and aHydro attractors being smaller than between the exact solution and third-order Chapman-Enskog hydrodynamics.

\begin{acknowledgments}
\vspace{-3mm}
The authors would like to express their gratitude to Michael McNelis for numerous clarifying discussions, and to the authors of Ref.~\cite{Behtash:2017wqg} for valuable comments. C.C. thanks Amaresh Jaiswal for insightful remarks. The work of UH and GV was supported in part by the U.S. Department of Energy (DOE), Office of Science, Office for Nuclear Physics under Award No. \rm{DE-SC0004286} and by the National Science Foundation (NSF) within the framework of the JETSCAPE Collaboration under Award No. ACI-1550223. GV also acknowledges support by the Fonds de Recherche du Qu\'ebec --- Nature et Technologies (FRQNT).
\end{acknowledgments}
\vfill


\end{document}